\documentclass[aps,twocolumn,english,superscriptaddress,citeautoscript,showkeys,preprintnumbers,amsmath,amssymb,floatfix,footinbib,prb]{revtex4-2}
\usepackage{xcolor}
\usepackage{soul}
\usepackage{amssymb,amsmath}
\usepackage{hyperref}
\hypersetup{colorlinks=true,linkcolor=red,citecolor=blue}
\usepackage[utf8]{inputenc}
\usepackage[english]{babel}
\usepackage{txfonts}

\usepackage{graphicx} % Include figure files
\usepackage{dcolumn} % Align table columns on decimal point
\usepackage{bm} % bold math
\usepackage{float}
\usepackage{hyperref}

\begin{document}
%\linenumbers

\title{Multigap Superconductivity in the Filled-Skutterudite Compound LaRu$_4$As$_{12}$ probed by muon spin rotation}

\author{A. Bhattacharyya}
\email{amitava.bhattacharyya@rkmvu.ac.in}
\affiliation{Department of Physics, Ramakrishna Mission Vivekananda Educational and Research Institute, Belur Math, Howrah 711202, West Bengal, India} 
\author{D. T. Adroja} 
\email{devashibhai.adroja@stfc.ac.uk}
\affiliation{ISIS Facility, Rutherford Appleton Laboratory, Chilton, Didcot, Oxon, OX11 0QX, United Kingdom} 
\affiliation{Highly Correlated Matter Research Group, Physics Department, University of Johannesburg, Auckland Park 2006, South Africa}
\author{M. M. Koza}
\email{koza@ill.fr}
\affiliation{Institut Laue Langevin, 71 avenue des Martyrs, CS20 156, 38042 Grenoble, Cedex 9, France.}
\author{S. Tsutsui}
\affiliation{Japan Synchrotron Radiation Research Institute (JASRI), SPring-8, Sayo, Hyogo 679-5198, Japan}
\affiliation{Institute of Quantum Beam Science, Graduate School of Science and Engineering,
Ibaraki University, Hitachi, Ibaraki 316-8511, Japan}
\author{T. Cichorek}
\affiliation{Institute of Low Temperature and Structure Research, Polish Academy of Sciences, 50-950 Wroclaw, Poland}
\author{A. D. Hillier}
\affiliation{ISIS Facility, Rutherford Appleton Laboratory, Chilton, Didcot, Oxon, OX11 0QX, United Kingdom}  

\begin{abstract}

Muon spin rotation ($\mu$SR) and inelastic X-ray scattering (IXS) were used to investigate the superconducting properties of the filled-skutterudite compound LaRu$_{4}$As$_{12}$. A two-gap isotropic ($s+s$)-wave model can explain the temperature dependence of the superfluid density. Zero field $\mu$SR measurements confirm that the time-reversal symmetry does not break upon entering the superconducting state. The measurements of lattice dynamics at 2, 20 and 300 K revealed temperature dependencies of the phonon modes that do not follow strictly a hardening of phonon frequencies upon cooling as expected within the quasi-harmonic picture. The 20~K data rather mark a turning point for the majority of the phonon frequencies.  Indeed a hardening is observed approaching 20~K from above, while for a few branches a weak softening is visible upon further cooling to 2~K. The observed dispersion relations of phonon modes throughout the Brillouin zone matches with the DFT prediction quite closely. Our results point out that cubic LaRu$_{4}$As$_{12}$ is a good reference material for studying multiband superconductivity, including those with lower crystallographic symmetries such as iron arsenide-based superconductors.

\end{abstract}

\date{\today} 


\maketitle

%----------------------------------------------------------------------------------------
	\section{Introduction}
%----------------------------------------------------------------------------------------

Materials scientists and physicists have been intrigued by properties of  the ternary transition metal pnictides with the chemical formula MT$_{4}$Pn$_{12}$ (M = alkali metal, alkaline-earth metal, lanthanide or light actinide element; T = Fe, Ru, or Os; Pn = P, As, or Sb), which crystallize in the filled skutterudite structure with space group $Im\bar{3}$ (No. 204)~\cite{Book1}. Examples include excellent thermoelectric performance ~\cite{Sales1996, Keppens1998}, and strongly correlated electron phenomena such as unconventional superconductivity ~\cite{EDBauer2002, Kotegawa2003, Adroja2005}, quadrupolar ordering ~\cite{Hachitani2006, Curnoe2002, Sugawara2002, Iwasa2002, Kohgi}, non-Fermi liquid behavior \cite{Takeda2000}, hybridization gap or Kondo phenomena~\cite{CeRu4Sb12_1,Takeda2000,Adroja2003,Adroja2007,Baumbach2008,Sanada2004} as well as unusual metal-insulator transitions ~\cite{Sekine}. From various factors that give rise to pletora of unuusal  physical properties of MT$_{4}$Pn$_{12}$, the hybridization of localized $f$-electron states with conduction electron states is of prime importance. As shown for, e.g., the [Ru$_{4}$As$_{12}$] sublattice, its subtle interplay with the M cation results in  multiband superconductivity, non-Fermi-liquid behavior, conventional superconductivity, and low-lying ferromagnetic order can be realized for La ~\cite{5}, Ce~\cite{6}, Pr ~\cite{7}, and Nd~\cite{8}, respectively. It is worth emphasizing that due to their cubic structure, MT$_{4}$Pn$_{12}$ compounds are unique in that they can be used to study multiband effects~\cite{Seyfarth,Jeitschko,Sato}. This opens an opportunity to compare and contrasts  multiband effects in the superconductivity of anisotropic materials, such as iron pnictides~\cite{Tarantini,Ren}, heavy-fermion compounds ~\cite{Seyfarth, Kasahara,Hill,Kittaka}, and
topological insulators ~\cite{Hor,Zhang}.

\begin{figure}[t]
\centering
\includegraphics[width=0.9\linewidth]{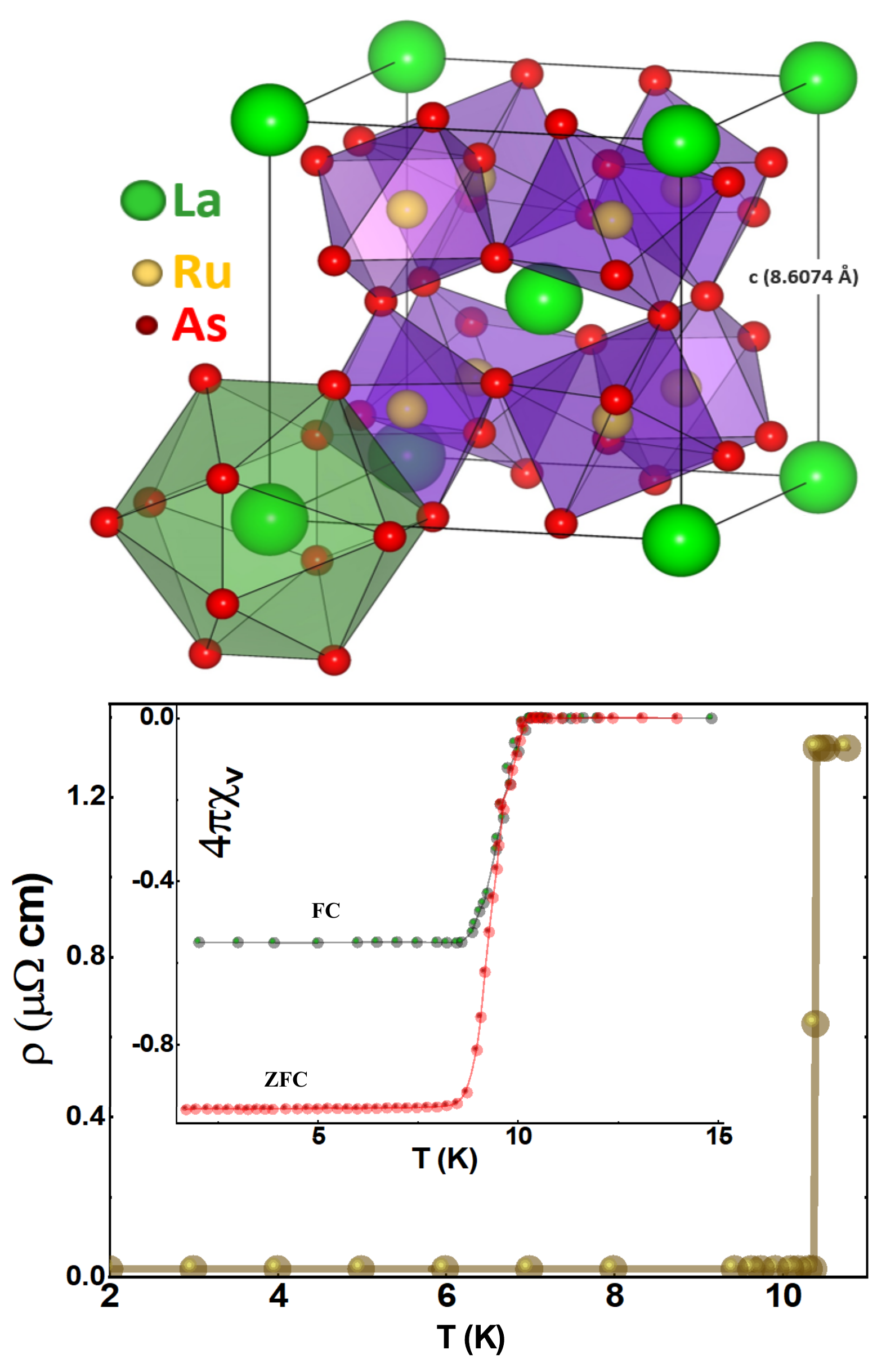}
\caption{(Top) The body-centered cubic structure of the filled skutterudite LaRu$_4$As$_{12}$ with space group $Im\bar{3}$. (bottom) The low-temperature electrical resisitivity of LaRu$_4$As$_{12}$ showing a sharp superconducting transition at 10.4 K. Inset: Zero-field-cooled (ZFC) and field-cooled (FC) magnetization in presence of 30 mT applied field as a function of temperature for a collection of LaRu$_{4}$As$_{12}$ single crystals.} 
\label{rt}
\end{figure}

\begin{figure}[t]
\centering
\includegraphics[width=\linewidth]{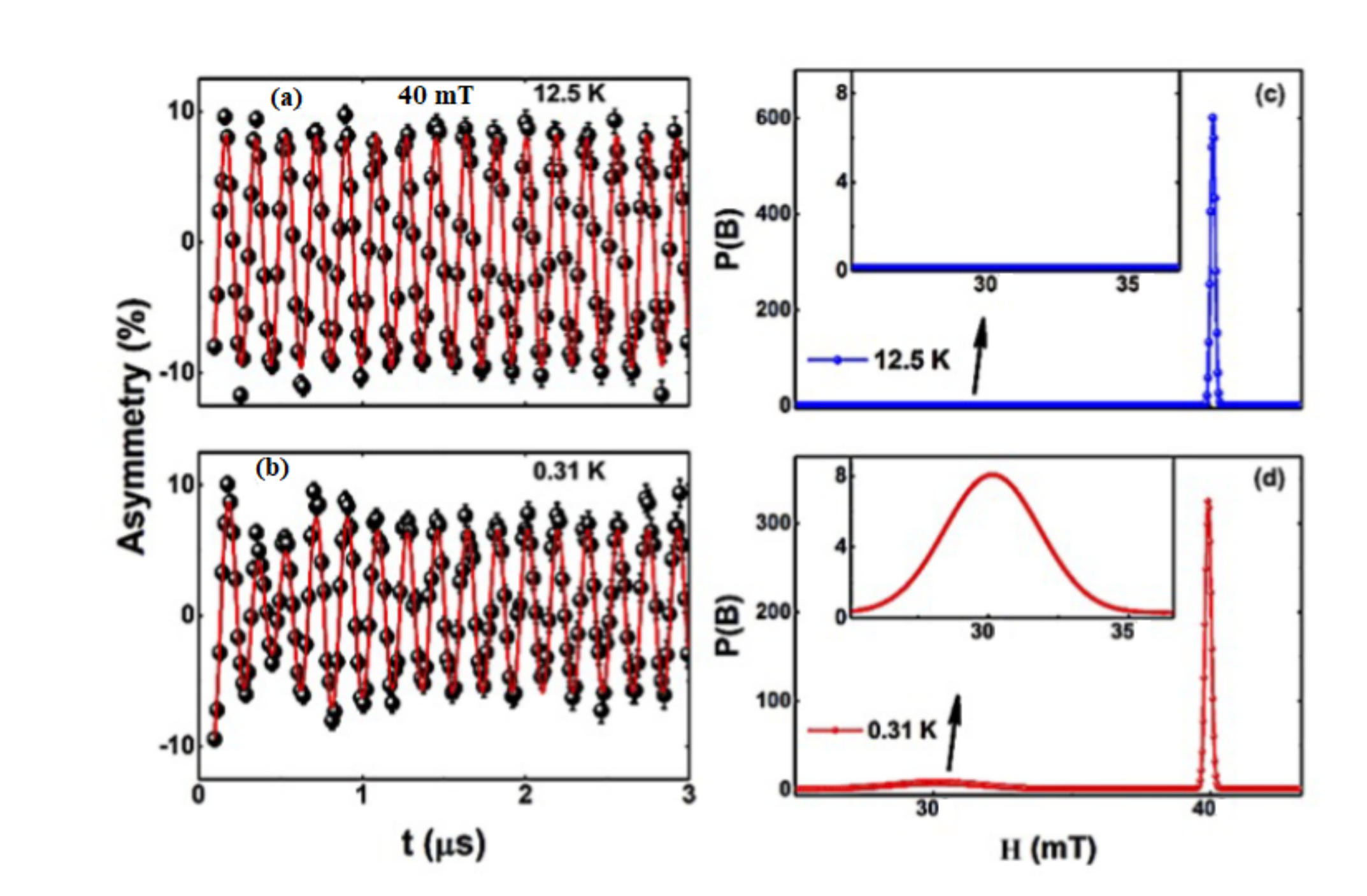}
\caption {(a)-(b) The asymmetry spectra of the TF-$\mu$SR data in the low time region, obtained in a 40 mT applied magnetic field at $T$ = 12.5 K (i.e. above $T_c$) and $T$ = 0.31 K (i.e. below $T_c$), respectively. (c)-(d) The maximum entropy plots at $T$ = 12.5 K and $T$ = 0.31 K, respectively. A zoomed-in section of the low-field region is shown in the inset.}
\label{tfmusr}
\end{figure}

\par

Most of the La-filled skutterudites exhibit BCS-type superconductivity with the superconducting critical temperature $T_c$ ranging from 0.4 K to 10.4 K~\cite{Jeitschko,Sato}. Among all completely filled compounds LaT$_4$Pn$_{12}$, the highest transition temperature accompanied by the highest upper critical field $H_{c2}$(0) = 10.2 T shows LaRu$_4$As$_{12}$~\cite{Namiki}. These parameters are substantially enhanced as compared to the sister compounds LaOs$_4$As$_{12}$~\cite{Shirotani} and PrRu$_4$As$_{12}$~\cite{Shirotani1997} with $T_c$  = 3.2 K and 2.3 K, respectively. This enhancement is even more intriguing, if one takes into account that LaRu$_4$As$_{12}$ does not show any distinct differences in both the conduction electron density of states at the Fermi level and the vibrational properties~\cite{Klotz}. Another remarkable feature of LaRu$_{4}$As$_{12}$ is several arguments for multiband superconductivity~\cite{Namiki,Klotz}. Specifically, the temperature dependence of the lower critical field $H_{c1}$ displays the sharp anomaly deep in the superconducting state, indicative of a rarely observed case of almost decoupled bands~\cite{Namiki}. Furthermore, a very recent observation of suppression of anharmonic phonons by artificial nonmagnetic defects is in line with a multiband superconductivity in the filled skutterudite LaRu$_{4}$As$_{12}$~\cite{Mizukami11}. Considering all of these factors, we are compelled to investigate this compound further.

\par

In this work, we investigated LaRu$_{4}$As$_{12}$ utilizing muon spin rotation and relaxation($\mu$SR), as well as inelastic X-ray scattering (IXS) measurements. $\mu$SR is a microscopic and very sensitive technique and provides direct information on the gap symmetry from the temperature dependence of superfluid density  measured using transverse field (TF) $\mu$SR.  Further zero-field (ZF) $\mu$SR is very sensitive to detect very small local magnetic field inside the (type II) superconducting state and hence can provide direct information on the  broken time reversal symmetry. Inelastic X-ray scattering provides direct information on phonon dispersion and hence it is very important to investigate change in the phonon dispersion above and below the superconducting transition in LaRu$_{4}$As$_{12}$. These measurements are important to understand the mechanism of the superconductivity in LaRu$_{4}$As$_{12}$. Using TF-$\mu$SR measurement, we show clear evidence of multiband superconductivity in LaRu$_4$As$_{12}$. As a result, in addition to Fe-based and cuprates superconductors, LaRu$_4$As$_{12}$ could be an unique multiband superconductor~\cite{Bochenek}. The preservation of time-reversal symmetry is revealed in ZF-$\mu$SR study. The observed phonon dispersion, using IXS, shows good agreement with the DFT-calculated phonon dispersion.

%----------------------------------------------------------------------------------------
\section{Experimental Details}
%----------------------------------------------------------------------------------------

Single crystals of LaRu$_4$As$_{12}$ were grown by mineralization in a molten Cd:As flux using a procedure previously reported~\cite{Henkie}. The low-temperature electrical resistivity of a LaRu$_{4}$As$_{12}$ single crystal was investigated by a conventional four-point ac technique  using a $^4$He cryostat (PPMS). Magnetic measurements for a collection of LaRu$_4$As$_{12}$ single crystals were performed down to 1.8 K at H = 30 mT utilizing a superconducting quantum interference device magnetometer (MPMS). The MUSR spectrometer at ISIS Pulsed Neutron and Muon Source, Rutherford Appleton Laboratory, UK, was used to measure the muon spin rotation and relaxation of LaRu$_4$As$_{12}$. 

%The $\mu$SR measurements are a valuable tool for determining the temperature variation of tiny field distribution within a type II superconductor, such as the spontaneous internal field arising from time reversal symmetry breaking phenomenon. 

The powder sample (1 g) of LaRu$_4$As$_{12}$ obtained from very small single crystals was mounted on a silver holder using diluted GE varnish (99.999\%). This was placed on the dilution refrigerator's cold finger. Zero-field  and transverse field measurements were taken at temperatures ranging from 0.31 K to 13.0 K. For ZF measurements, an active correction coil was employed to cancel any stray magnetic fields at the sample space to a level of 10$^{-4}$ mT. 
The ZF-$\mu$SR measurement assists in identifying the spontaneous internal field associated with time-reversal symmetry breaking~\cite{Sonier}. The TF-$\mu$SR measurements were carried out in the presence of a 40 mT external magnetic field, which is much higher than the lower critical field ($\mu_0H_{c1}$ = 1.3 mT) and significantly lower than the upper critical field ($\mu_0H_{c2}$ = 10.3 T) of LaRu$_4$As$_{12}$. To avoid any flux trapping effects, we applied the transverse field much above $T_{c}$ and subsequently cooled the sample down to base temperature. The WiMDA software was used to evaluate the experimental data~\cite{Pratt2000}.

\begin{figure}[t]
\centering
\includegraphics[width=0.9\linewidth]{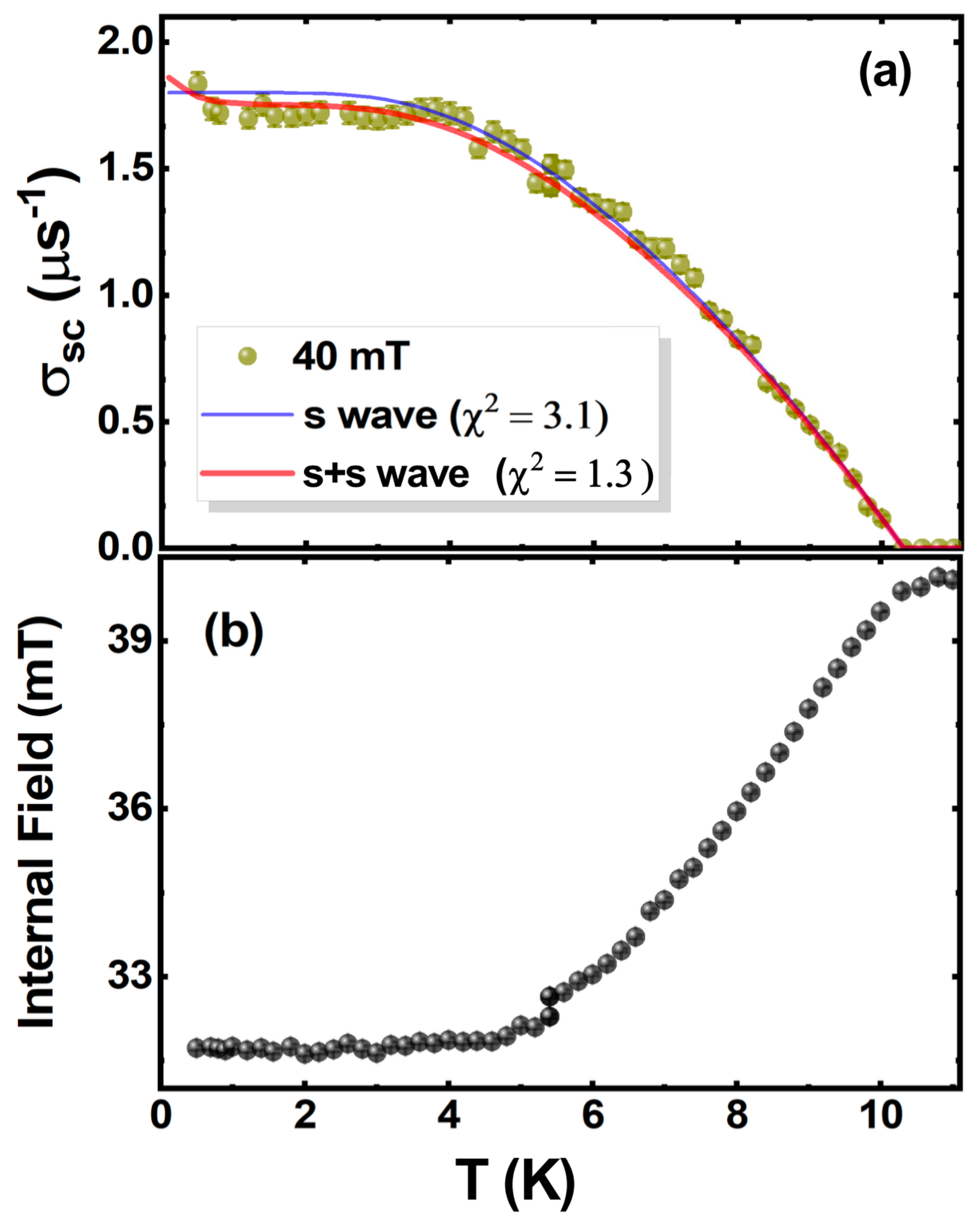}
\caption {(a) The temperature dependence of $\sigma_{sc}(T)$ data with fits with various gap models. The red solid line shows the fit to the two-gap $s$+$s$-wave model, while the blue line shows the fit to the single-gap isotropic $s$-wave model.  (b) Temperature variations in the internal field confirm the appearance of Meissner state.}
\label{tfmusr2}
\end{figure}

\par
\begin{figure}[b]
\centering
\includegraphics[width=0.9\linewidth]{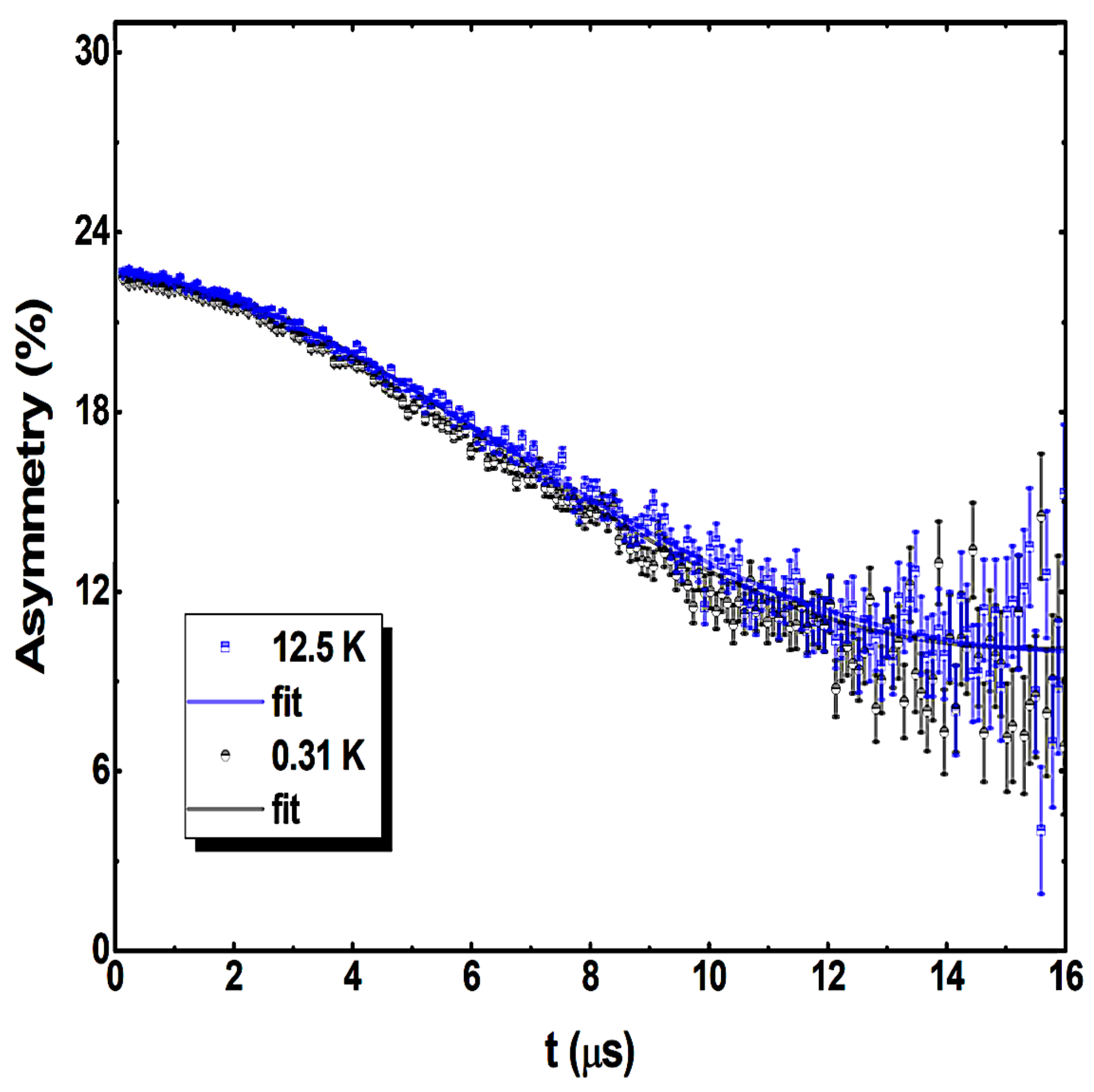}
\caption{The ZF-$\mu$SR spectra for LaRu$_4$As$_{12}$ at 0.31 K (black) and 12.5 K (blue). As stated in the text, the solid line fits the data.}
\label{zfmusr}
\end{figure}

\par

High-resolution inelastic X-ray scattering (IXS) experiments were carried out at the beamline BL35XU of the SPring-8 synchrotron radiation facility in Harima, Japan~\cite{BL35XU-Baron-JPCS-2000}. Si(11 11 11) backscattering optics with 21.747 keV photon energy was chosen for the present work resulting in an energy resolution of 1.5 meV. The $Q$-resolution was set to (0.06 0.06 0.06) in reciprocal lattice units along [1 0 0] and [1 1 0] directions. Constant $Q$ scans were carried out in an extended range of reciprocal space at temperatures 2, 20, and 300~K. Few supplementary scans were performed at 6 and 14~K. i.e below and above $T_ c$.  The sample temperature was controlled by a helium closed-cycle refrigerator. The quality of the single-crystal sample was checked by measuring the rocking curve widths on a series of Bragg reflections along with high-symmetry directions, say [1 0 0], [1 1 0], and [0 0 1]. These widths are less than 0.1 degrees and close to the resolution limit of the BL35XU spectrometer, $\sim$ 0.02 degrees. Phonon properties of LaRu$_4$As$_{12}$ were modelled by density functional theory (DFT) based lattice dynamics calculations. The first-principles calculations were performed with the Vienna ab-initio simulation package (VASP) utilizing projector augmented wave potentials and the generalized gradient approximation of Perdew-Burke-Ernzerhof (GGA-PBE) for the exchange-correlation terms~\cite{VASP-Kresse-1996,PAW-Kresse-1999,GGA-PBE-PRL-1996}. The electronic minimization, as well as the subsequent calculation of Hellmann-Feynman (HF) forces, were carried out with a 2x2x2 supercell and a Monkhorst-Pack k-mesh of 4x4x4~\cite{Monkhorst-PRB-1976}. The first-order Methfessel-Paxton method with a $\sigma=0.1$ for the band occupancies was applied and forces were relaxed to 10$^{-7}$. The relaxed structure resulted in a lattice parameter of a = 8.6074~\AA~ and the fractional coordinates of $y = 0.3497$, $z = 0.1507$ for the As position. The dynamical matrix and, thus, the phonon eigenvectors and eigenstates were derived from the HF forces by the direct method implemented in the software package PHONON~\cite{Phonon-Parlinski-1999}. The HF forces were computed for symmetry non-equivalent displacements of 0.03~\AA~ in high symmetry directions of the atoms. For the estimation of phonon intensities, the eigenvectors were scaled by the atom-specific squared electron number $Z^2$ and an averaged electronic form-factor. Within the dynamic range given by the experimental setup, the highest phonon intensities were derived for $Q$ numbers around the (600) Bragg reflection and the high symmetry directions [600]$\rightarrow$[611],[600]$\rightarrow$[700]$\rightarrow$[6.5 .5 .5].

%----------------------------------------------------------------------------------------
\section{Results and discussion}
\subsection{Structure \& Physical Properties}
%----------------------------------------------------------------------------------------

LaRu$_4$As$_{12}$ has a body-centered cubic structure with space group $Im\bar{3}$ (No. 204) which crystallizes in the CoAs$_3$-type skutterudite structure packed with La atoms. Green are La, yellow are Ru, and As are red atoms, as seen in Fig.~\ref{rt}(a). The electropositive element La, which lacks four-fold rotational symmetry, is at the centre of the large icosahedron cage formed by As atoms. Ru, a transition metal ion, creates a primitive cubic sublattice between the cages. The atomic positions are (0, 0, 0) for 1 La atom in the crystallographic position (2$a$), (1/4, 1/4, 1/4) for 4 Ru atoms in (8$c$), (0, $y$, $z$) for 12 As atoms in (24$g$), with $y$ = 0.3499 and $z$ = 0.1502. Fig.~\ref{rt} (b) displays the temperature variation of resistivity in the zero applied magnetic field in low temperature limit. The electrical resistivity reveals an on set of superconductivity at $T_c$ = 10.4 K. The low residual resistivity data revealed the good quality of the sample. The superconducting volume fraction in the LaRu$_4$As$_{12}$ sample was calculated using magnetization measurements. The inset of Fig.~\ref{rt}(b) shows the magnetization data for both zero field cooled (ZFC) and field cooled (FC) measurements. In the ZFC measurement, full diamagnetic shielding is observed below $T_c$ = 10.4 K. The upper critical field is estimated to be 10.3 T, exceeding the Pauli paramagnetic limit 18.4 $T_c$ = 18.9 T.

\begin{figure}[t]
\begin{center}
\includegraphics[width = \linewidth, height = \linewidth]{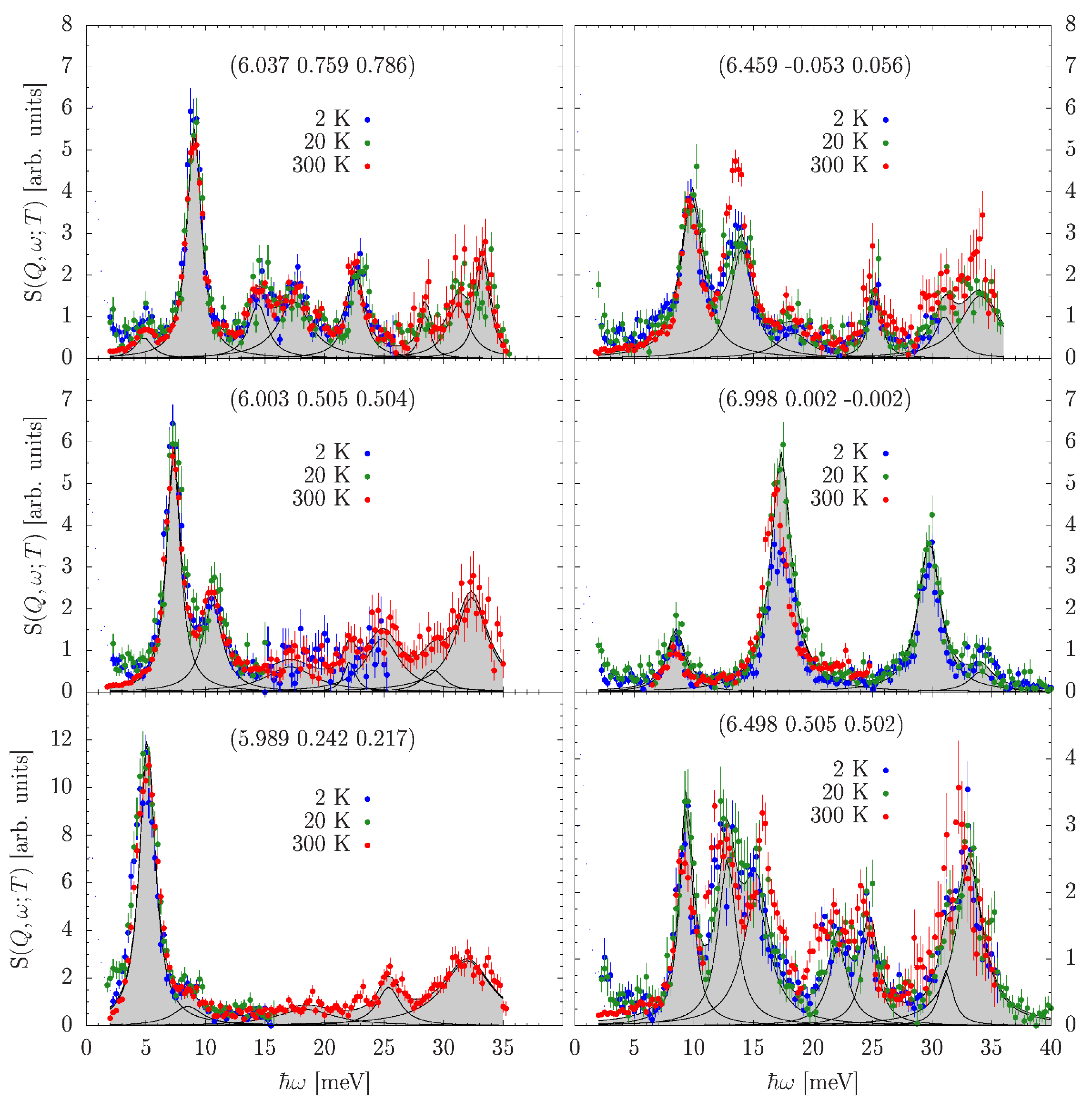}
\caption{\textcolor{black}{Selected spectra of LaRu$_4$As$_{12}$ recorded along high-symmetry directions at the three different $T$ indicated in the figures. Lines with data correspond to standard Lorentzians fitted to the data (300~K in left panel and 20~K in right panel). Filled areas highlight the total fit functions.}}
\label{fig_IXS_spectra}
\end{center}
\end{figure}

%----------------------------------------------------------------------------------------
\subsection{Superconducting gap structure: TF-$\mu$SR}
%----------------------------------------------------------------------------------------

To further understand the multiband nature of superconductivity as suggested by the earlier experimental~\cite{Juraszek2020} and theoretical studies~\cite{Klotz}, we have performed TF-$\mu$SR measurements in LaRu$_4$As$_{12}$. The TF-$\mu$SR muon spin precession signals collected at $T$ = 12.5 K and 0.31 K in $H$ = 40 mT are shown in Figs.~\ref{tfmusr}(a)-(b). Due to the flux-line lattice state, which causes spatial inhomogeneity of the magnetic field distribution below $T_c$, the data in Figs.~\ref{tfmusr}(a)-(b) indicate a distinct contrast in the relaxation rate above and below $T_c$. The maximum entropy plots in Figs.~\ref{tfmusr}(c)-(d) show one peak at 12.5 K at the applied field. An extra peak at 0.31 K is observed on the lower side of the applied field in addition to the peak at 40 mT, confirming type-II superconductivity in this sample. Use of straight Fourier transformation to extract frequency spectra from transverse $\mu$SR data is not practicable, since at long times the data are noisy since the count rate is low, due to the shortness of the muon lifetime.  On the other hand the maximum entropy method overcomes this problem and allows simultaneous analysis of spectra from multiple detectors with different phases to yield a single frequency spectrum \cite{Rainford}. The inset of  Fig.~\ref{tfmusr}(d) shows a zoomed section of the low-field region of the maximum entropy plot, which reveals a broad peak near 30 mT associated with the vertex lattice formation.\\

\par

Due to the single peak of the internal field in the superconducting state, we have fitted the data using a Gaussian function having one component. To illustrate the characteristics of the vortex state, we have fitted the data using the expression~\cite{Bhattacharyya1, Bhattacharyyarev,ThCoC2,Panda}, 
\begin{equation}
A_\mathrm{TF}(t) = A_\mathrm{0}\cos(\gamma_\mathrm{\mu}H_0t+\Phi)\exp(-\frac{\sigma^{2}t^{2}}{2})+A_\mathrm{bg}\cos(\gamma_\mathrm{\mu}H_{bg}t+\Phi)
\end{equation}
where the first component is the sample contribution and the second is the background contribution. The initial asymmetries of the sample and silver holder contributions are described by $A_0$ and $A_{bg}$, respectively, where $A_{bg}$ does not undergo any depolarization; the internal fields of the sample and sample holder are described by $H_0$ and $H_{bg}$, respectively. The muon gyromagnetic ratio is $\gamma_\mathrm{\mu}$/2$\pi$ = 135.53 MHz/T; $\sigma$ is the Gaussian muon spin relaxation rate; $\Phi$ is the initial phase of the signal at the detectors. The values of $A_{0}$=0.66(1) and $A_{bg}$=0.34(1) were estimated by fitting 0.31 K data and which were kept fixed during the analysis of other temperature data. The superconducting depolarization rate can be determined by subtracting the nuclear contribution: $\sigma_\mathrm{sc} = \sqrt{\sigma^{2}-\sigma_\mathrm{n}^2}$, where $\sigma_{\mathrm{n}}$ = 0.047 $\mu$s$^{-1}$, is the normal state contribution, which is temperature independent as evident in spectra above $T_c$. Since $\sigma_\mathrm{sc}$ is connected to the magnetic penetration depth ($\lambda$) by $\sigma_\mathrm{sc}~\approx~1/\lambda^2$ for a triangular lattice~\cite{Chia,Sonier}, the temperature variation of $\sigma_\mathrm{sc}(T)$ can be used to measure the nature of the superconducting gap. Then, using the functional form given below, a single- and a two-gap $s$-wave model is fitted to the data:
\begin{align}
\label{MuonFit3}
\frac{\sigma_{sc}(T)}{\sigma_{sc}(0)} &= \frac{\lambda^{-2}(T)}{\lambda^{-2}(0)} \nonumber  \\ 
&= w\frac{\lambda^{-2}(T,\Delta_{0,1})}{\lambda^{-2}(0,\Delta_{0,1})}+(1-w)\frac{\lambda^{-2}(T,\Delta_{0,2})}{\lambda^{-2}(0,\Delta_{0,2})}
\end{align}
where $\lambda(0)$ is the value of the penetration depth at T = 0 K, $\Delta_{0,i}$ is the value of the i$^{th}$ (i = 1 or 2) superconducting gap at T = 0 K and $w$ is the weighting factor of the first gap. Each term in Eq. (2) is evaluated using the standard expression within the local London approximation ($\lambda$ $>>$ $\xi$ ) ~\cite{Prozorov} as
\begin{align}
\label{MuonFit2}
\frac{\sigma_{\mathrm{sc}}\left(T\right)}{\sigma_{\mathrm{sc}}\left(0\right)} &= \frac{\lambda^{-2}\left(T,\Delta_{0,i}\right)}{\lambda^{-2}\left(0,\Delta_{0,i}\right)} \nonumber \\
 &=1+\frac{1}{\pi}\int_{0}^{2\pi}\int_{\Delta\left(T,\phi\right)}^{\infty}\left(\frac{\partial f}{\partial E}\right) \frac{E\mathrm{d}E\mathrm{d}\phi}{\sqrt{E^{2}-\Delta_i^2\left(T,\phi_{}\right)}},
\end{align}

where $f= [1+\exp(E/k_\mathrm{B}T)]^{-1}$ is the Fermi function, the temperature and  angle-dependent gap function is expected to follow the relationship: $\Delta_{i}(T,\phi) = \Delta_{0,i}\delta(T/T_\mathrm{C})\mathrm{g}(\phi)$. Here $\Delta_{0,i}$ denotes the maximum gap value at zero temperature and $\mathrm{g}(\phi)$ is the angular dependence of the gap, $\mathrm{g}(\phi)$ = 1 for an isotropic $s$-wave model and for $s+s$ wave gaps. Here $\phi$ is the azimuthal angle. The temperature depedence of the superconducting gap symmetry is expected to follow, $\delta(T/T_\mathrm{C}) = \tanh[1.82[1.018(T_\mathrm{C}/T-1)]^{0.51}]$\cite{Annett,Pang}. At any coupling strength, this gap function is accurate enough to explain the temperature dependency.\\~\\

\begin{table}
\caption{Fitted parameters obtained from the fit to the$\sigma_\mathrm{sc}$(T) data of LaRu$_{4}$As$_{12}$ using different gap models.}
\vspace{0.5cm}
\resizebox{\linewidth}{!}{
    \begin  {tabular}{|c|c|c|c|c| }
      \hline
    {Model} &   {$\Delta_{i}(0)$(meV)} &{2$\Delta_{i}(0)$/k$_\mathrm{B}T_\mathrm{C}$} &w& $\chi^{2}$ \\ \hline
        s-wave &  1.59 & 3.57& 1 & 3.1\\ 
        
       s+s wave              & 1.656, 0.064       & 3.73, 0.144 & 0.87& 1.3 \\
        \hline
   \end{tabular}}

\label{Tabel}
\end{table}

\par

\begin{figure*}[t]
\begin{center}
\includegraphics[width = \linewidth]{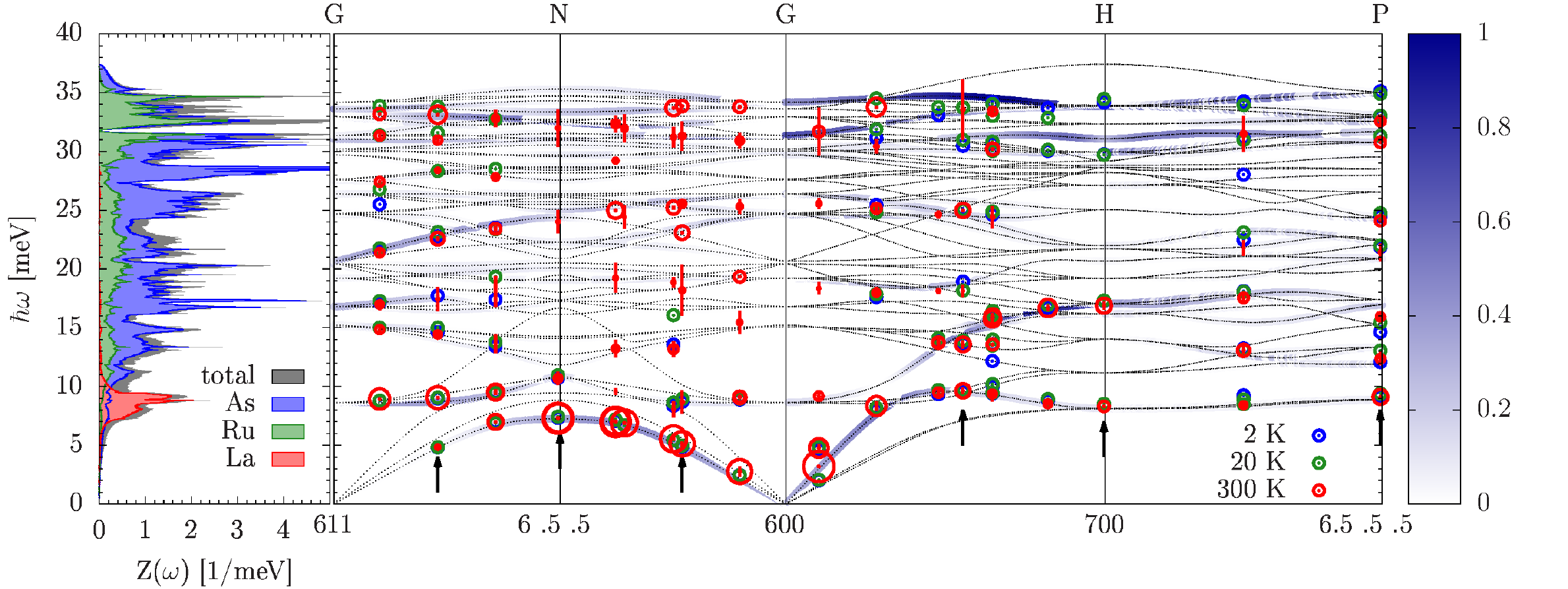}
\caption{\textcolor{black}{Left, total and partial phonon densities of states of LaRu$_4$As$_{12}$ computed with DFT and normalized to 51 phonon modes in the total signal. Right, phonon energies derived from the Lorentzian ﬁts. Symbol sizes correspond to the logarithm of the peak intensities, and error bars indicate the full width at half maximum of the peaks. Dotted lines with the data report the DFT computed phonon dispersion relation. Blue shaded signal indicates the DFT-computed phonon intensities. The energy scale of both, the Z$(\omega)$ and the phonon dispersion, has been scaled up by 7~\% for a better match with the experimental data.}
\label{fig_DFT_dispersion}}
\end{center}
\end{figure*}

\begin{figure}[b]
\begin{center}
\includegraphics[width = \linewidth]{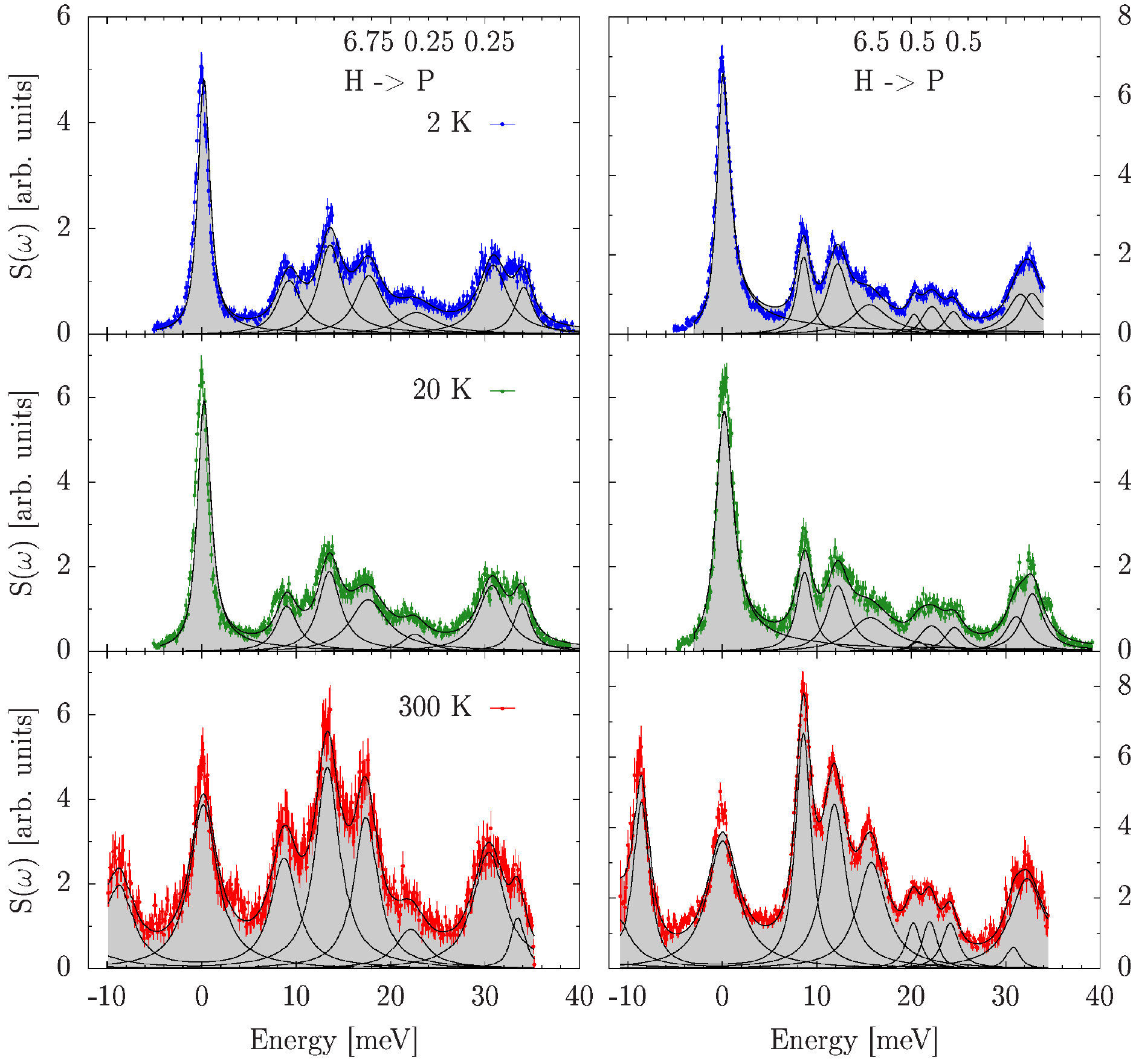}
\caption{\textcolor{black}{Generalized spectra S$(\omega)$ at $T=2, 20$ and $300$~K from top to bottom.
Left and right, data from scans for high symmetry mid-point and end-point scan in direction H$\rightarrow$P.
Lines with the data correspond to standard Lorentzians fitted to the data. Gray shaded area highlights the fitted total signal.}
\label{fig_generalized_spectra}}
\end{center}
\end{figure}

\begin{figure}[h]
\begin{center}
\includegraphics[width = \linewidth]{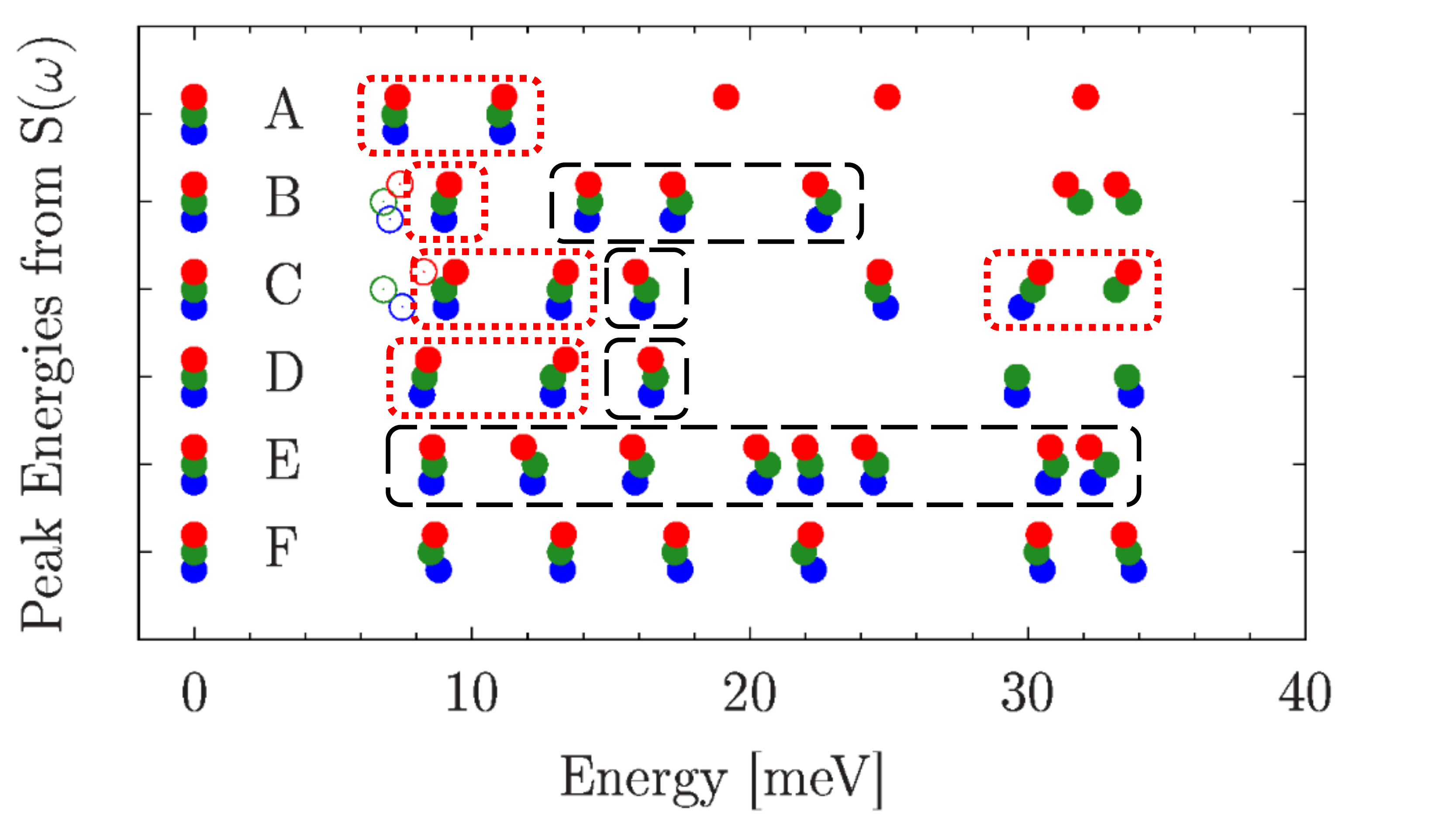}
\caption{\textcolor{black}{Temperature dependence of peak energies derived through Lorentzian fits to the generalized S$(\omega)$ spectra.
Different $T$ are disclosed by the colors as blue (2~K), green (20~K), and red (300~K).
Characters indicate the different setups applied with
A ($\Gamma\rightarrow$ N, 600 $\rightarrow$ 6 .5 .5), 
B ($\Gamma\rightarrow$ N, 6.5.5 $\rightarrow$ 611), 
C ($\Gamma\rightarrow$ H, 600 $\rightarrow$ 6.500), 
D ($\Gamma\rightarrow$ H, 6.500 $\rightarrow$ 700), 
E (H $\rightarrow$ P, 6.5.5.5),
and F (H $\rightarrow$ P, 6.75.25.25)).
Results for E and F are highlighted in Fig.~\ref{fig_generalized_spectra} and for all other orientations in Fig.~8 of the Supplementary Information.
Boxes framed by dashed black lines highlight frequencies at which a turnover from mode hardening to softening upon cooling is particularly obvious.
Boxes framed by dotted red lines indicate frequencies at which a softening is observed below 300~K not in line with a standard quasi--harmonic response.
See text for details. The symbol size approximates a diameter of 1 meV. The corresponding reliability parameters of the fits are reported in Table I of the supplementary information section. They are of the order of 0.1 meV.} 
\label{fig_IXS_energypoints}}
\end{center}
\end{figure}

\par

The symmetry of the superconducting gap was determined by fitting the data of $\lambda^{-2}(T)$ or $\sigma_{sc}(T)$ using (a) a single gap $s$-wave and (b) a multigap $s$+$s$-wave gap model, as presented in Fig.~\ref{tfmusr2}(a). Because it produces a high $\chi^2$ goodness of fit value (here $\chi^2=\sum{(\sigma_{obs}-\sigma_{cal})^2}/{\sigma_{error}^2(M-N)}$, the total number of data points is M, and the number of free parameters is N), the typical BCS single gap cannot characterise the $\lambda^{-2}(T)$ data. A recent study of LaRu$_4$As$_{12}$ by Juraszek {\it et al.}~\cite{{Juraszek2020}} confirms that the lower critical field behavior is well suited to multiband isotropic $s$-wave superconductivity. The multiband character of LaRu$_4$As$_{12}$ is further strengthened by the two-gap model employing the isotropic $s$+$s$-wave with $\chi^2$ = 1.3, compared to $\chi^2$ = 3.1 for single gap $s$-wave fit. The estimated larger gap is 2$\Delta_1(0)/k_BT_C$ = 3.73$\pm$0.2 in the $s$+$s$-wave model, which is close to the value of 3.53 driven from BCS theory, and the estimated smaller gap is 2$\Delta_2(0)/k_BT_C$ = 0.144$\pm$0.01 in the $s$+$s$-wave model. The smaller value of the second gap is observed in many Fe-based superconductors, for example RbCa$_{2}$Fe$_{4}$As$_{4}$F$_{2}$ $2\Delta_{1}/k_{B}T_{C}$ = 6.48 and $2\Delta_{2}/k_{B}T_{C}$= 0.7 $T_{C}$ = 29.2 K) [see Table-I of Ref.~\cite{Bhattacharyyarev}]. The different size of the gaps is arising from the Fermi surface topology. Multigap features are common in iron-based superconductors e.g. Ba$_{1-x}$K$_{x}$Fe$_{2}$As$_{2}$~\cite{Khasanov2009,Avci}, cuprate superconductors~\cite{Khasanov2007} also in Bi$_{4}$O$_{4}$S$_{3}$~\cite{Biswas}. We calculated the values of London penetration depth $\lambda_L(0)$= 240(4) nm for $s$+$s$ wave fit, carriers density $n_\mathrm{s} = 8.6(3) \times 10^{27}$ carriers m$^{-3}$, and effective mass of the quasiparticle $m^{*} = 1.749(2) m_\mathrm{e}$ for LaRu$_4$As$_{12}$ using the method described in Ref.~\cite{HfIrSi,CeIr3,Zr5Pt3}. As shown in Fig.~\ref{tfmusr2}(b), the internal field of the flux line lattice is temperature dependent and decreases with decreasing temperature, as in other superconductors, indicating diamagnetic shift. Because of the formation of a flux line lattice in the superconducting state, the total internal field inside the sample is reduced.

%----------------------------------------------------------------------------------------
\subsection{Time reversal symmetry: ZF-$\mu$SR}
%----------------------------------------------------------------------------------------

Here we present our ZF-$\mu$SR results for detecting the spontaneous internal field associated with time reversal symmetry breaking in the superconducting state. 
%By detecting the spontaneous internal field below the superconducting transition, zero field $\mu$SR directly probes the unconventional nature of the superconducting pairing mechanism. 
As shown in Fig.~\ref{zfmusr}, we measured the time evolution of ZF asymmetry spectra in both the normal and superconducting states. The ZF-$\mu$SR spectra were fitted using a combination of Lorentzian and Gaussian Kubo-Toyabe relaxation functions~\cite{ThCoC2}:
\begin{equation}
A_\mathrm{ZF}(t) = A_\mathrm{2}G_\mathrm{KT}(t)e^{-\lambda_\mathrm{ZF}t}+A_\mathrm{bg}
\end{equation}
here,
\begin{equation}
G_\mathrm{KT}(t) = [\frac{1}{3}+\frac{2}{3}(1-\sigma_\mathrm{KT}^{2}t^{2})\exp({-\frac{\sigma_\mathrm{KT}^2t^2}{2}})]
\end{equation}
is the Gaussian Kubo-Toyabe function, the sample and silver holder asymmetry contributions are represented by $A_\mathrm{2}$ and $A_\mathrm{bg}$, respectively.  As shown in Fig.~\ref{zfmusr}, both asymmetry spectra fall on top of each other, indicating that there is no change in relaxation rate in the superconducting state compared to the normal state. This denotes the lack of a spontaneous internal field. As a result, time reversal symmetry is preserved in the superconducting state. Furthermore, the fits to the ZF data yield, $\sigma_\mathrm{KT} = 0.11(1) ~\mu \mathrm{s}^{-1}$ and $\lambda_\mathrm{\mu} = 0.010(2) ~\mu \mathrm{s}^{-1}$ at 0.31~K and $\sigma_\mathrm{KT} = 0.105(3) ~\mu \mathrm{s}^{-1}$ and $\lambda_\mathrm{\mu} = 0.011(2) ~\mu \mathrm{s}^{-1}$ at 12.5~K.

%----------------------------------------------------------------------------------------
\subsection{Inelastic X-Ray Scattering}
%----------------------------------------------------------------------------------------

As outlined above, we have investigated the lattice dynamics of the multiband superconductor LaRu$_4$As$_{12}$ ($T_c$ =10.3 K) by IXS at 300 K, 20 K, and 2 K.
All spectra recorded at those $T$ are reported in Figs.~1--7 of the Supplementary Information. 
For the highest intensities, the measurements were focused near the (6, 0, 0) Bragg position by performing longitudinal and transverse scans. Fig.~\ref{fig_IXS_spectra} shows some selected spectra recorded along high-symmetry directions at the sampled temperatures of 300 K, 20 K, and 2 K.
It can be seen that the overall behavior of the compound is rather harmonic. No strong phonon renormalization can be identified over the entire explored temperature range. This applies not only to the high-symmetry directions shown in Fig.~\ref{fig_IXS_spectra} but also to any other direction examined.

All spectra have been fitted after the temperature correction with the Bose occupation number.
Experimentally determined resolution functions have been convoluted with the fitted model.
The model was formed from a set of standard Lorentzian functions whose number was adjusted to the number of peaks identified. To compare data recorded at the three different temperatures the final fits were carried out to the Stokes-line data only. However, we confirmed that both side fits of the spectra at 300~K resulted in very similar parameters. Differences might be attributed to the strong elastic line in the prior data treatment, which has been suppressed in the final fits.

Fig.~\ref{fig_DFT_dispersion} reports all derived phonon energies along high-symmetry directions compared with our DFT computed dispersion relation and the total and partial phonon densities of states Z$(\omega)$. Note, that for best match, the energy scale of the DFT data has been scaled up by 7~\%. This underestimation of eigenfrequencies and overestimation of structural dimensions is a well-known feature of the GGA-PBE and has been reported before~\cite{Koza-JPSJ-2013, Tutuncu-PRB-2017}.   The size of the energy symbols reflects the measured peak intensities on a logarithmic scale, and the peak widths are indicated as error bars. Vertical arrows indicate the ${\bf Q}$ points at which the spectra shown in Fig.~\ref{fig_IXS_spectra} have been recorded. 

Improving the statistical significance of the spectra was necessary to derive a conclusion upon the temperature dependence of the phonon modes.
S$({\bf Q},\omega)$ spectra from different analyzers recorded at a fixed setup were corrected for zero shift of the elastic line and merged to a generalized S$(\omega)$ signal.
An example of the resulting data is presented in Fig.~\ref{fig_generalized_spectra}.
All spectra have been fitted with standard Lorentzians accounting for the resolution function by convolution as well as the Bose thermal occupation. 
Fig.~\ref{fig_IXS_energypoints} reports the frequencies derived from those fits.
Different temperatures monitored at a single setup are disclosed by different colors.
Open symbols indicate less reliable frequencies which are strongly biased by the presents of the elastic line on one hand and a strong localized mode on the other.   

Frequencies for which a switchover from hardening (going from 300~K to 20~K) to softening upon cooling below T$_C$ (going from 20~K to 2~K) is observed are highlighted by the dashed black line boxes, see Fig.~\ref{fig_IXS_energypoints}.
Some other frequencies showing a softening already below 300~K are framed by dotted red line boxes.
The mode hardening between 300 and 20~K can be associated with the lattice contraction and is comprehensible within the theory of quasi-harmonic crystals.
Clearly, the mode softening below T$_C$ does not follow this classical quasi-harmonic concept. 
It rather points at a weak electron-phonon coupling effect when entering the superconducting state below ~10.5~K.
This observation is supported by the theoretically calculated $T_c$ = 11.56~K based on the electron-phonon coupling compared to the measured $T_c$= 10.45 K for LaRu$_4$As$_{12}$ ~\cite{Tutuncu2017}.

Note that the significance of the conclusion upon the detectability of the electron-phonon coupling in the present IXS data is qualified by the statistics of the results.
Further experiments are required to apprehend if specific vibrational eigenmodes are particularly involved in the electron-phonon interaction.
A participation of specific eigenmodes would result in a characteristic directional dependence of the phonon response and should be studied in future experiments.

%----------------------------------------------------------------------------------------
\section{CONCLUSION}
%----------------------------------------------------------------------------------------

In conclusion, we have determined the superconducting state of LaRu$_4$As$_{12}$ using TF-$\mu$SR and its lattice dynamics has been studied by inelastic X-ray scattering. The temperature dependence of magnetic penetration depth has been found to be consistent with a two-gap $s$+$s$-wave model of multiband superconductivity. The larger gap to $T_c$ ratio calculated within the $s$+$s$-wave scenario, 2$\Delta_1(0)/k_{\mathrm{B}}T_c$ = 3.73$\pm0.2$, is very close to the value of 3.53 for an ordinary BCS superconductor. The ZF-$\mu$SR spectra at 0.31 K and 12.5 K resemble each other very closely, indicative that the time-reversal symmetry is preserved in the superconducting  state of LaRu$_4$As$_{12}$.
Furthermore, IXS analyses of the Brillouin region reveal phonon modes between 300 and 20 K that show a weak temperature dependence comprehensible as an effect of the crystal's volume changes. 
Comparatively to 20 K, many of the 2 K modes show a weak softening of the phonon frequencies indicating that the electron-phonon interactions driving the
the superconductivity of LaRu$_4$As$_{12}$ exert a visible effect on the vibrational eigenstates. 
This observation is also in  agreement with the theoretically calculated $T_c$ = 11.56 K based on the electron-phonon coupling compared to the measured $T_c$= 10.45 K ~\cite{Tutuncu2017}. From a broader perspective our findings should be relevant for a wider class of multiband superconductors, including these with lower crystallographic symmetries such as iron-arsenide superconductors.\\

%----------------------------------------------------------------------------------------
\section{Acknowledgements}
%----------------------------------------------------------------------------------------

We are thankful to Prof. Tanmoy Das and Priyo Adhikari for their insightful discussion on superconducting gap structure. We are grateful to Prof. Zygmunt Henkie for providing single crystals of LaRu$_4$As$_{12}$. AB thanks the Science \& Engineering Research Board for the CRG Research Grant (CRG/2020/000698). The IXS experiment was performed under the approval of JASRI 
(Proposal No. 2018B1493).  DTA would like to acknowledge the funding from the Royal Society of London for Newton Advanced Fellowship between UK and China and for the International Exchange between UK and Japan and EPSRC-UK (Grant reference EP/W00562X/1).

\newpage \newpage 
%----------------------------------------------------------------------------------------
\section{Supplementary Information}
%----------------------------------------------------------------------------------------

\begin{figure*}[]
\begin{center}
\includegraphics[angle=0,width=125mm]{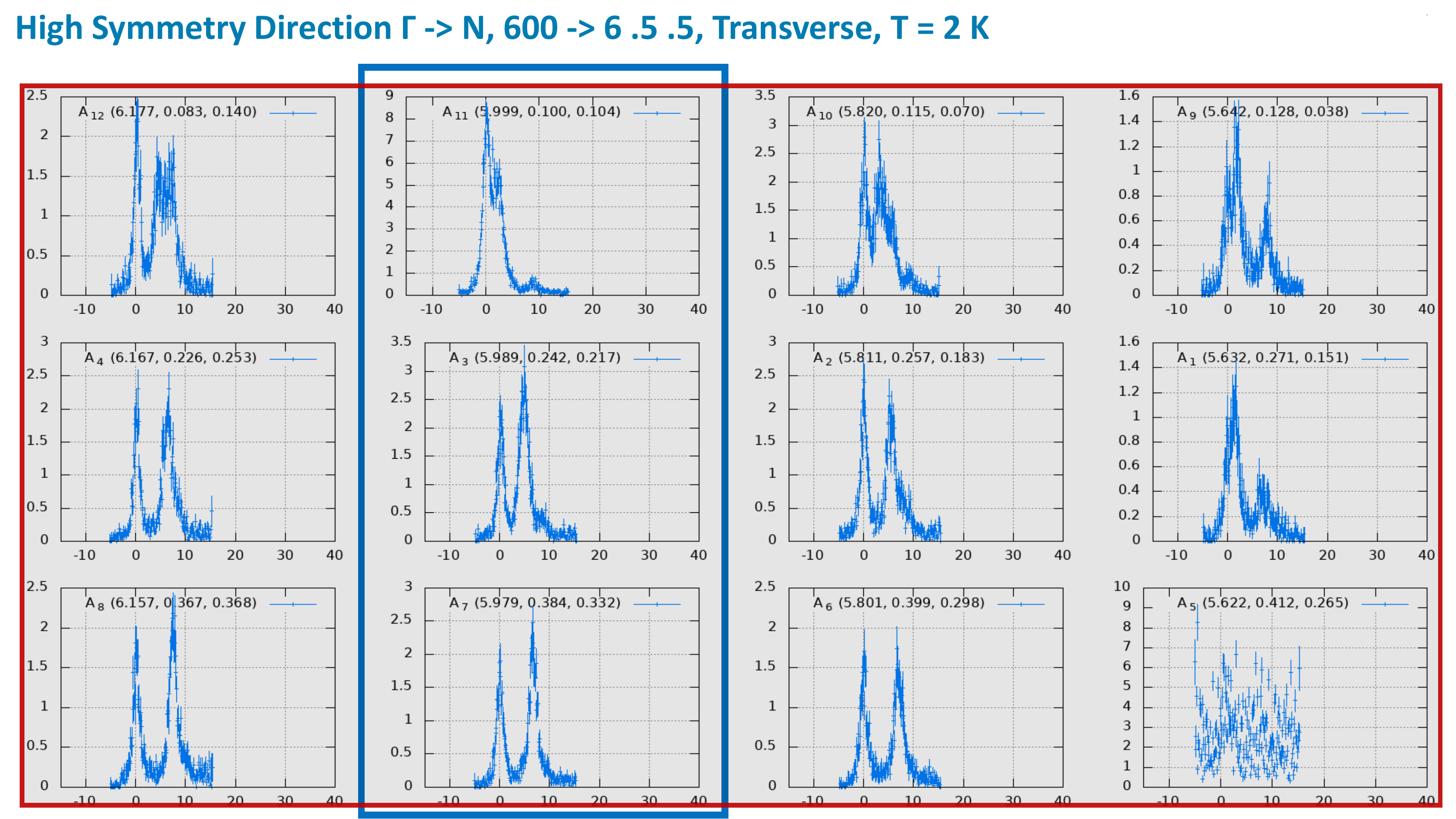}
\includegraphics[angle=0,width=125mm]{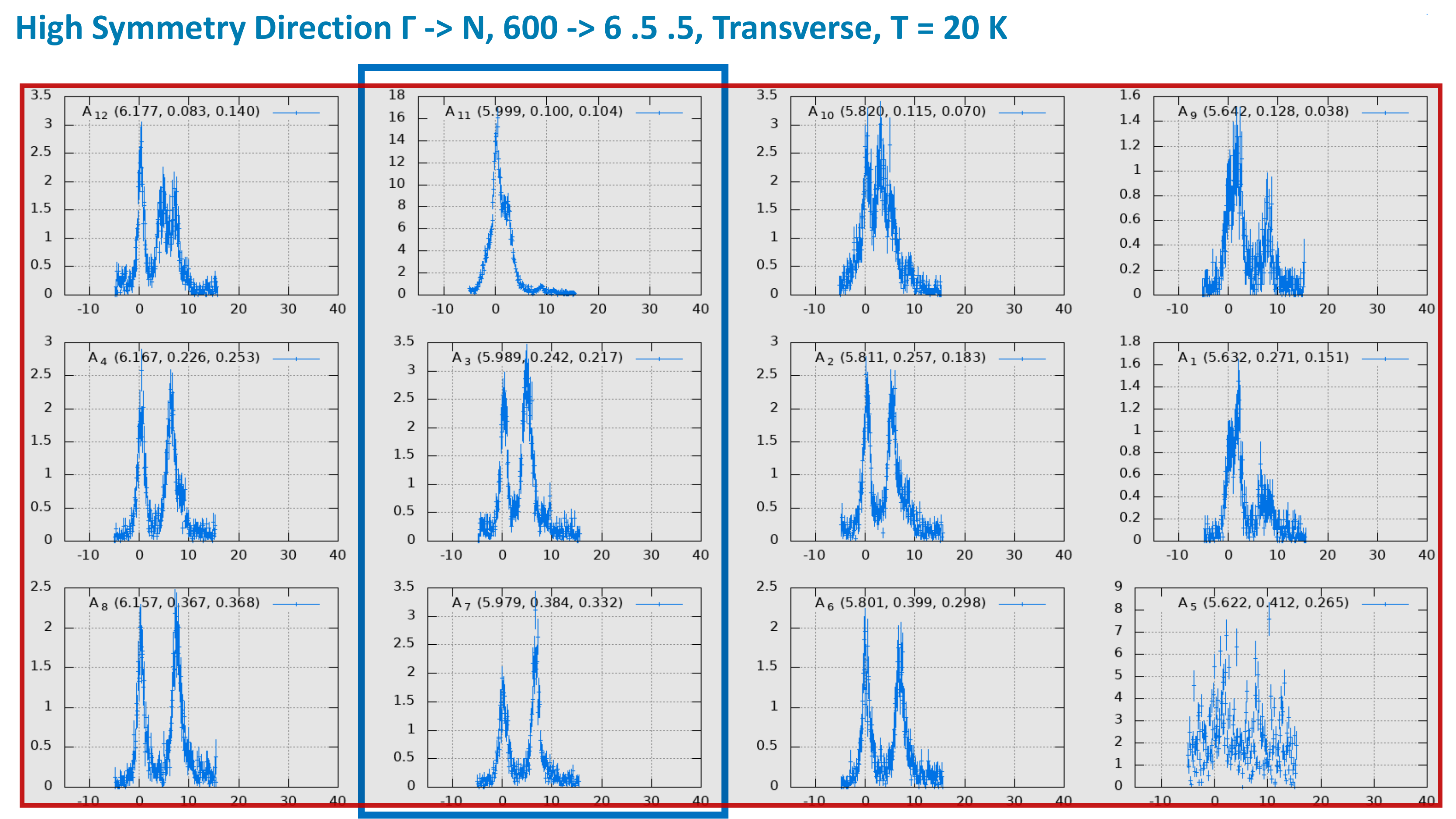}
\includegraphics[angle=0,width=125mm]{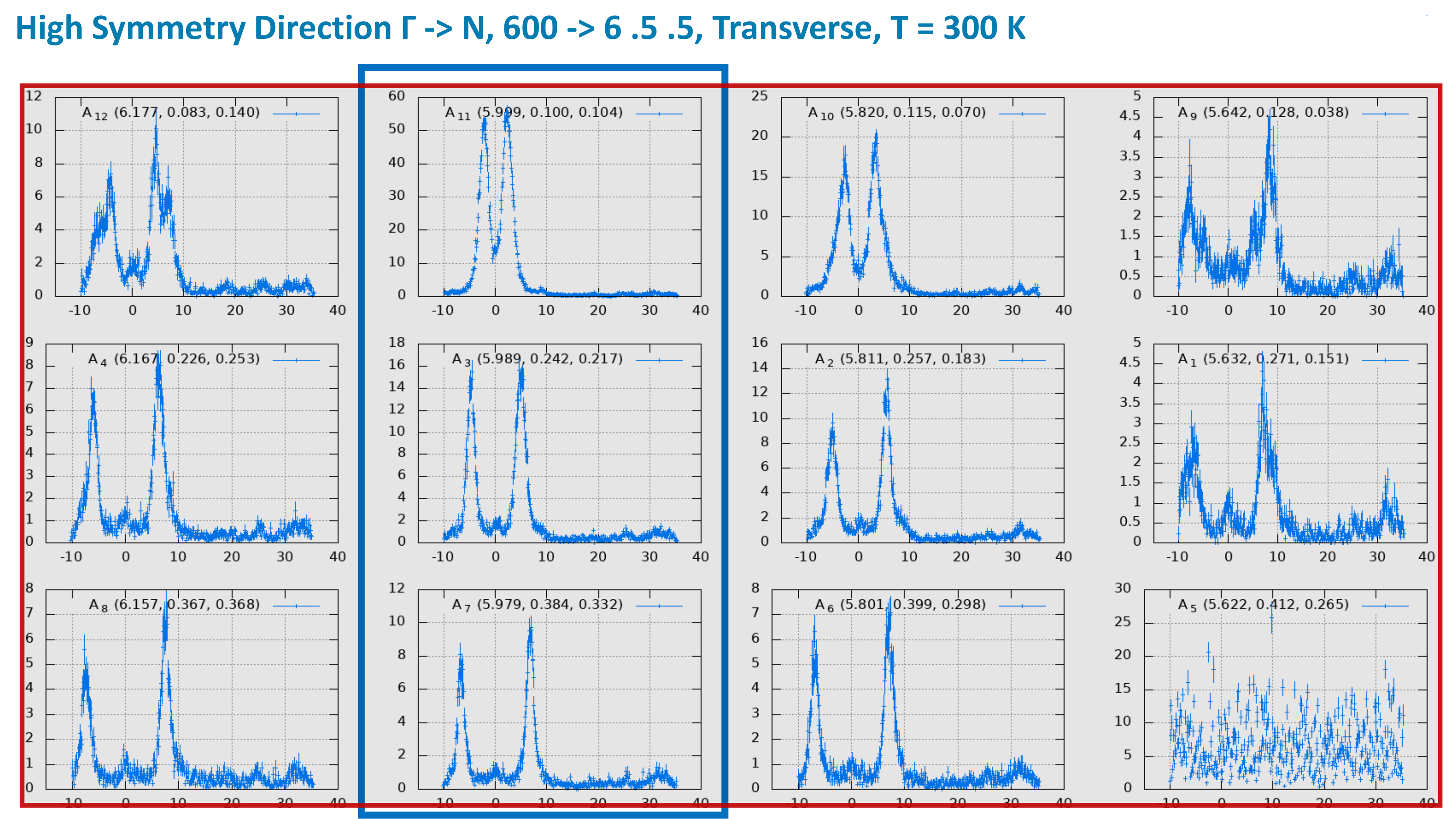}
\caption{Recorded spectra at 2, 20, and 300~K from top to bottom.
Constant \textbf{$Q$} scans were set up for high symmetry direction G$\rightarrow$N and transverse polarisaton of acoustic phonons highlighted by the blue-framed data set.
\textbf{$Q$} numbers are reported in the figures.
Red-framed spectra were not considered for the generalized S$(\omega)$ analysis.
\label{fig_SI_GNT1}}
\end{center}
\end{figure*}

\begin{figure*}[]
\begin{center}
\includegraphics[angle=0,width=125mm]{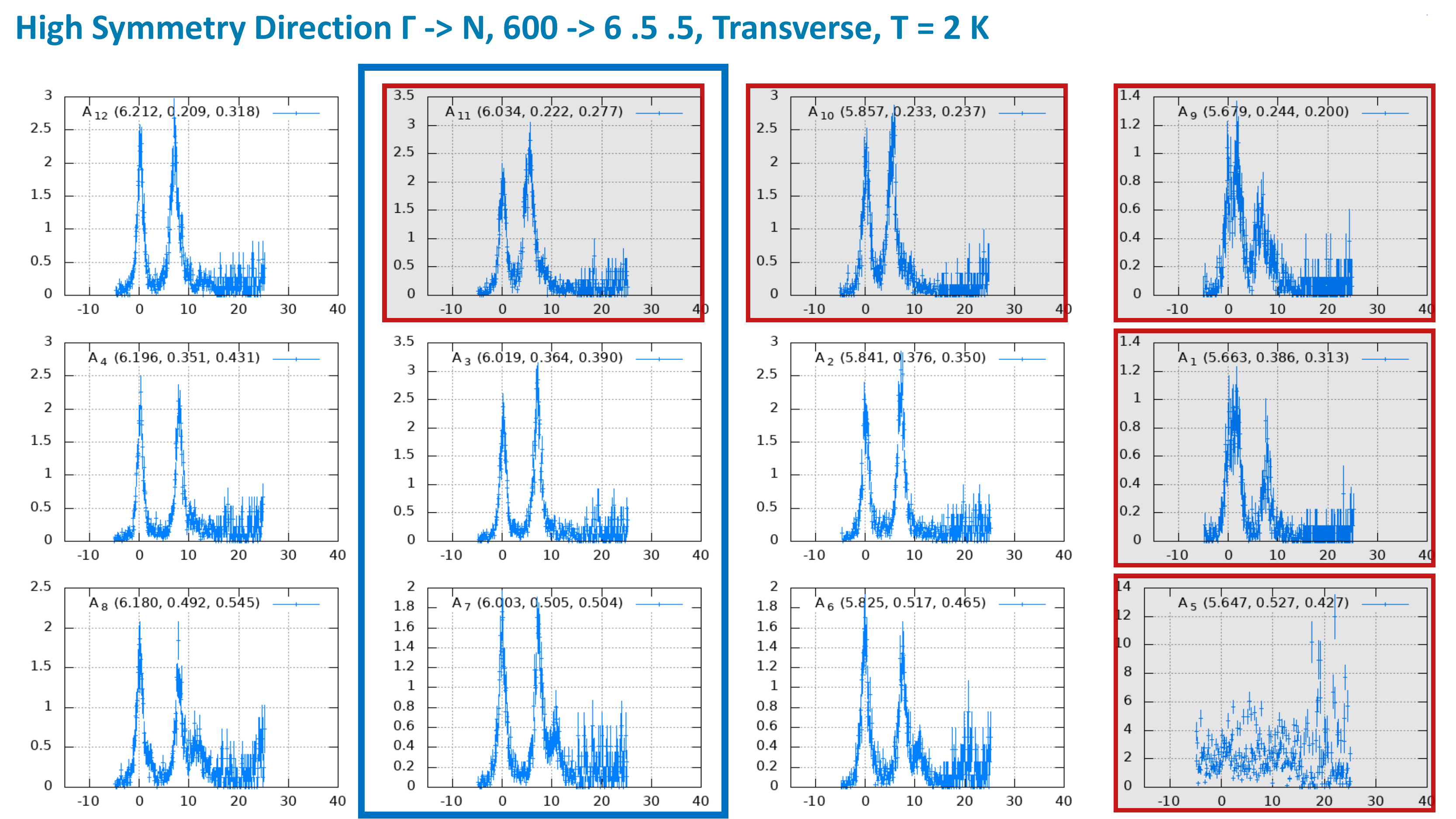}
\includegraphics[angle=0,width=125mm]{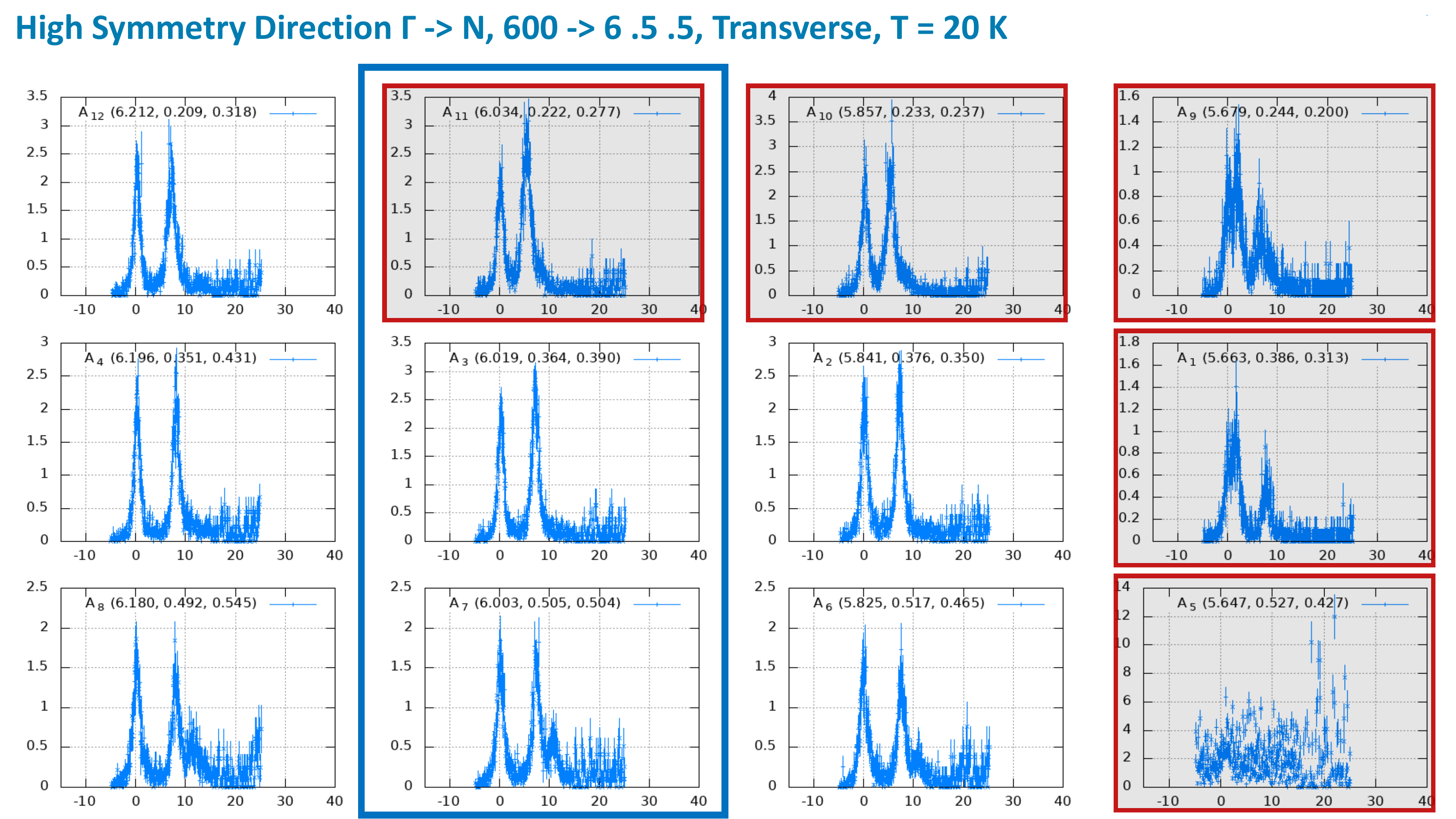}
\includegraphics[angle=0,width=125mm]{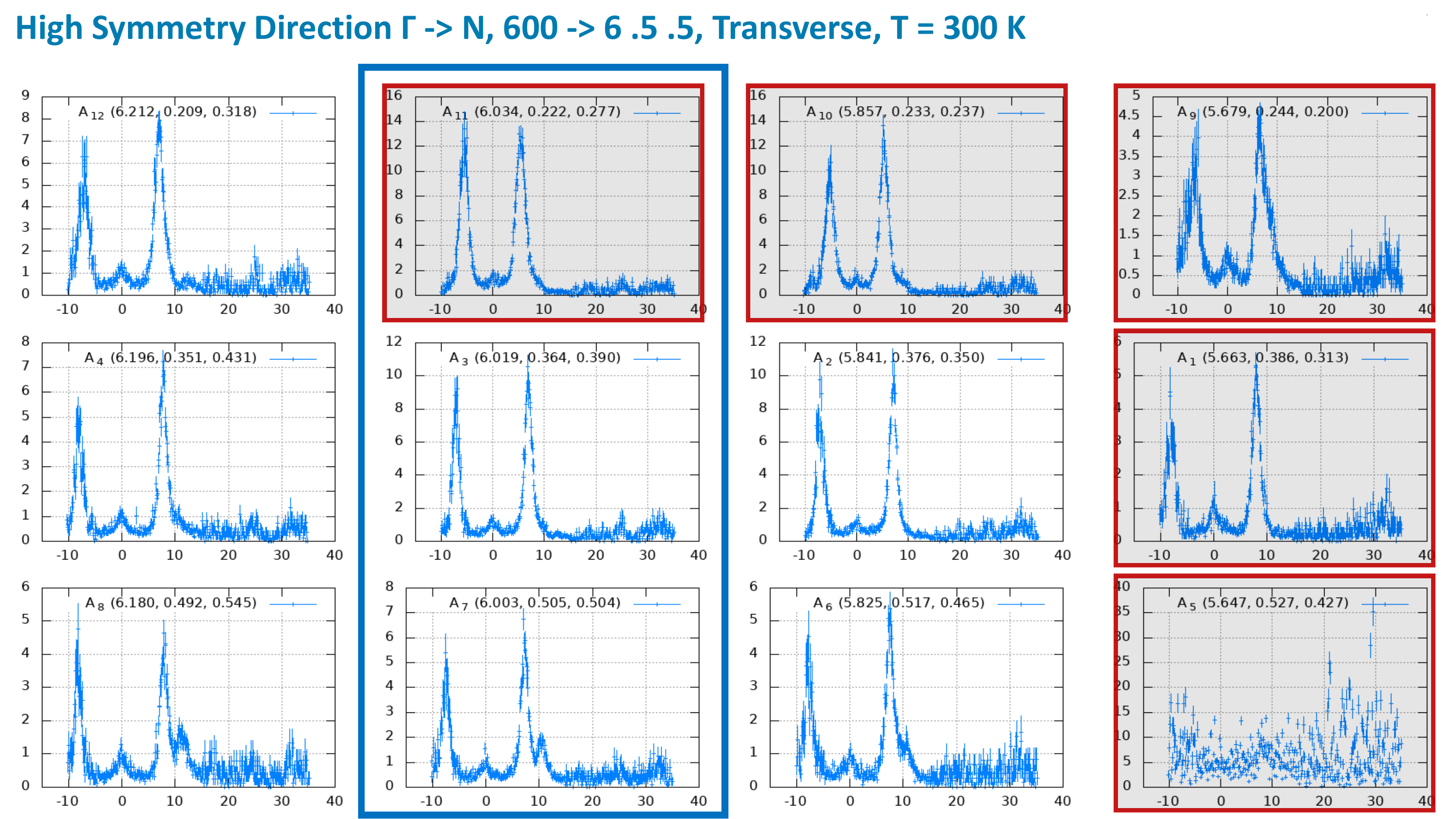}
\caption{Recorded spectra at 2, 20, and 300~K from top to bottom.
Constant \textbf{$Q$} scans were set up for high symmetry direction G$\rightarrow$N and transverse polarisaton of acoustic phonons highlighted by the blue-framed data set.
\textbf{$Q$} numbers are reported in the figures.
Red-framed spectra were not considered for the generalized S$(\omega)$ analysis.
\label{fig_SI_GNT2}}
\end{center}
\end{figure*}

\begin{figure*}[]
\begin{center}
\includegraphics[angle=0,width=125mm]{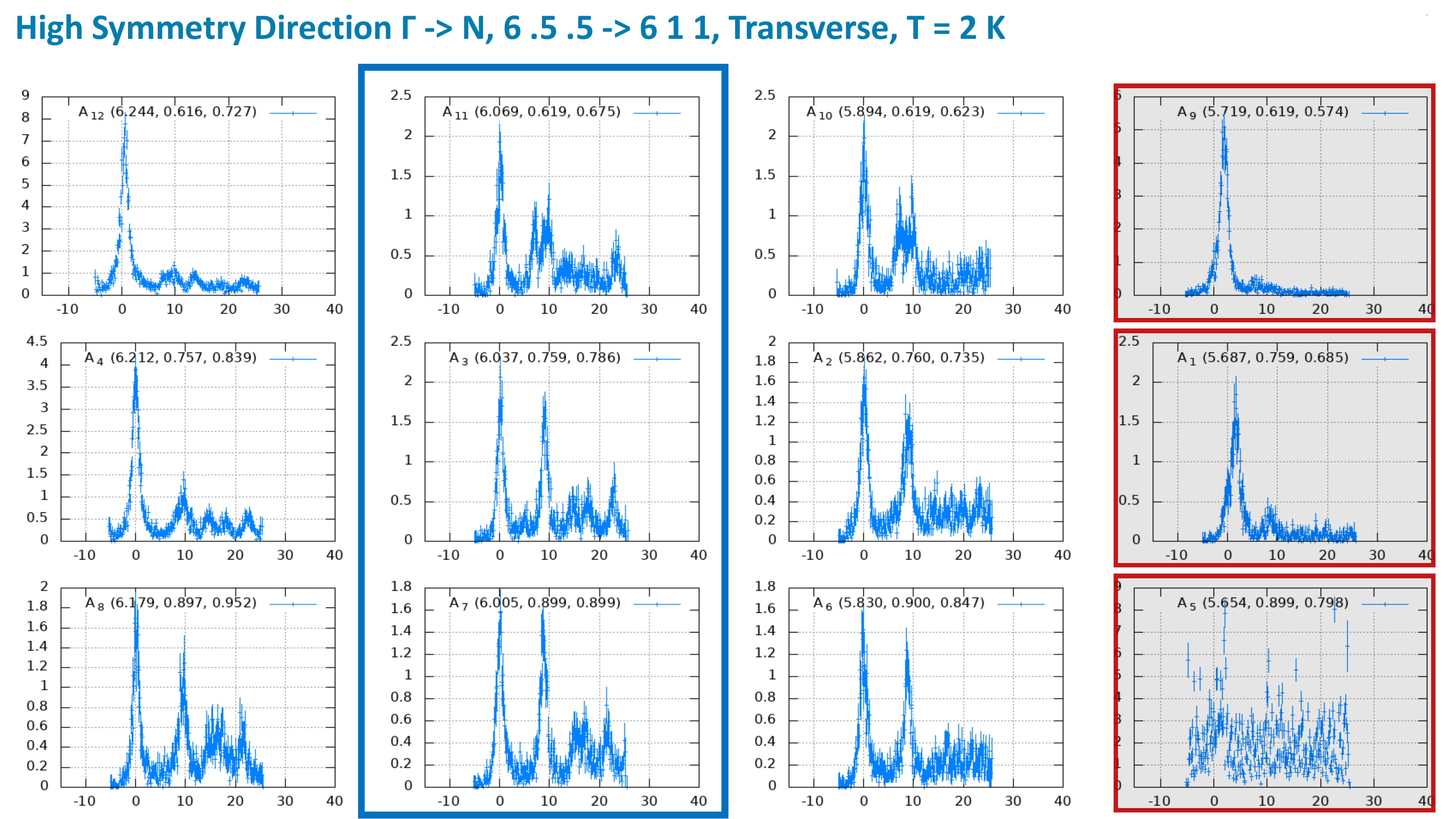}
\includegraphics[angle=0,width=125mm]{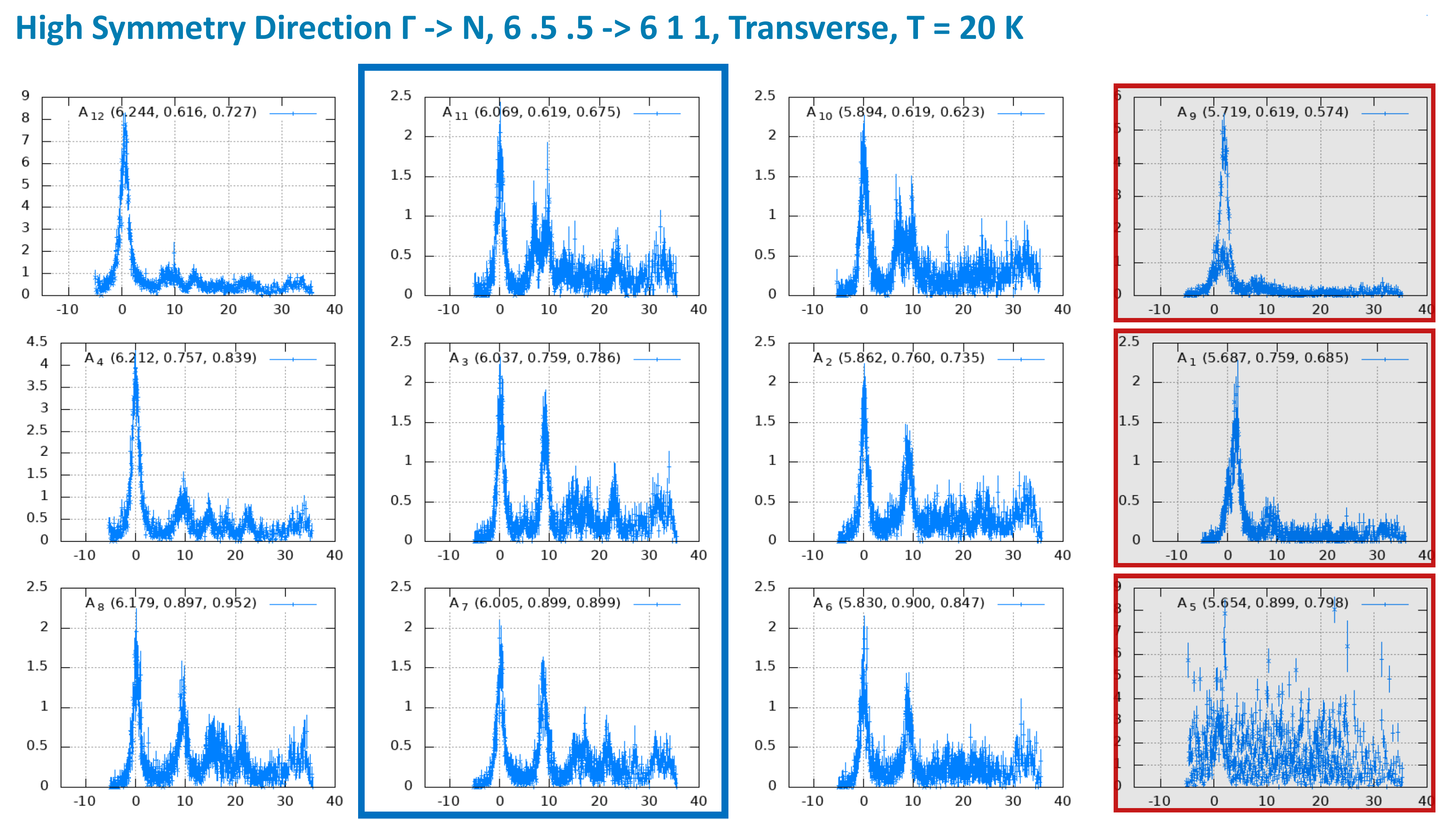}
\includegraphics[angle=0,width=125mm]{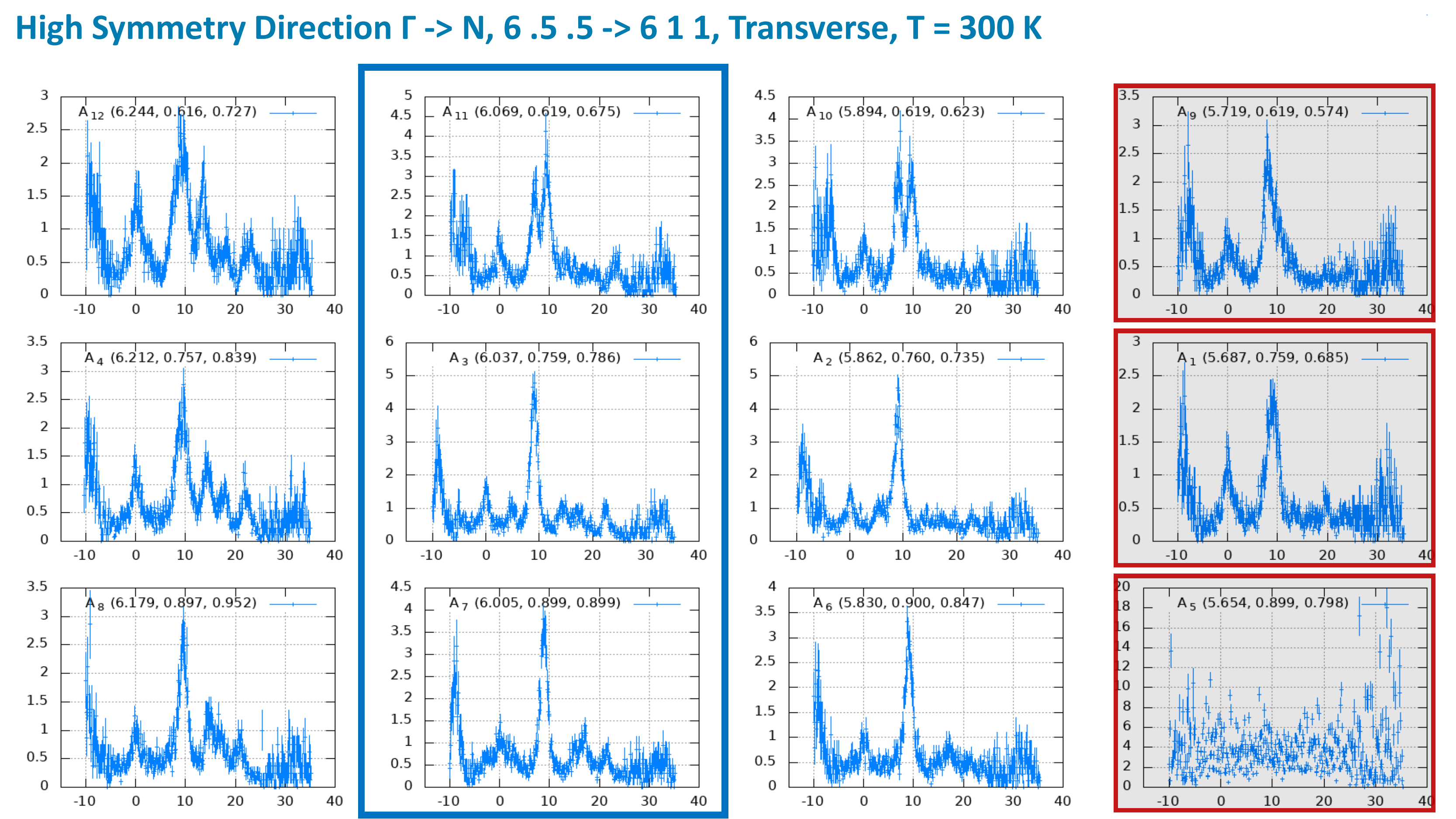}
\caption{Recorded spectra at 2, 20, and 300~K from top to bottom.
Constant \textbf{$Q$} scans were set up for high symmetry direction G$\rightarrow$N and transverse polarisaton of acoustic phonons highlighted by the blue-framed data set.
\textbf{$Q$} numbers are reported in the figures.
Red-framed spectra were not considered for the generalized S$(\omega)$ analysis.
\label{fig_SI_GNT3}}
\end{center}
\end{figure*}

\begin{figure*}[]
\begin{center}
\includegraphics[angle=0,width=125mm]{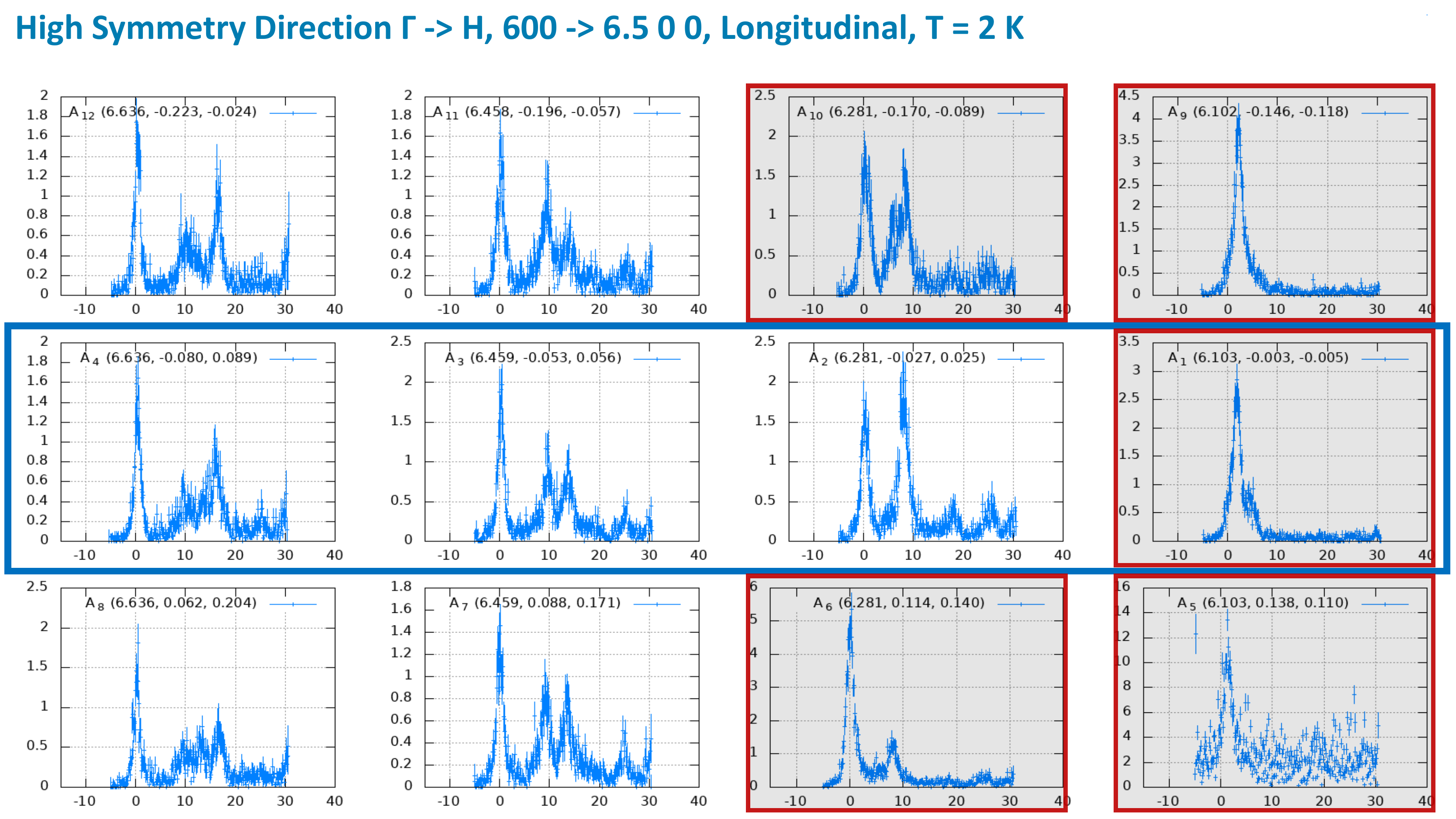}
\includegraphics[angle=0,width=125mm]{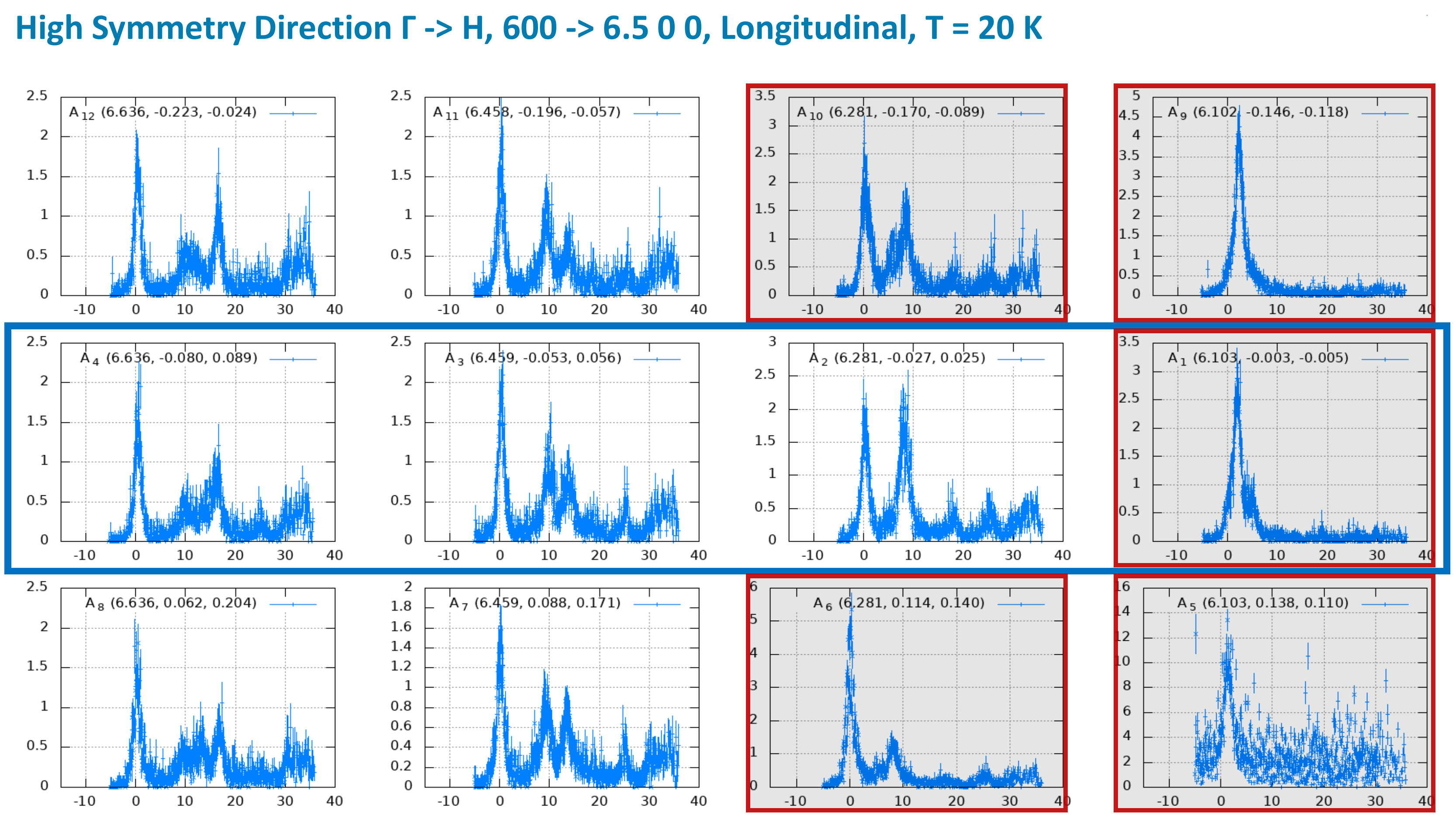}
\includegraphics[angle=0,width=125mm]{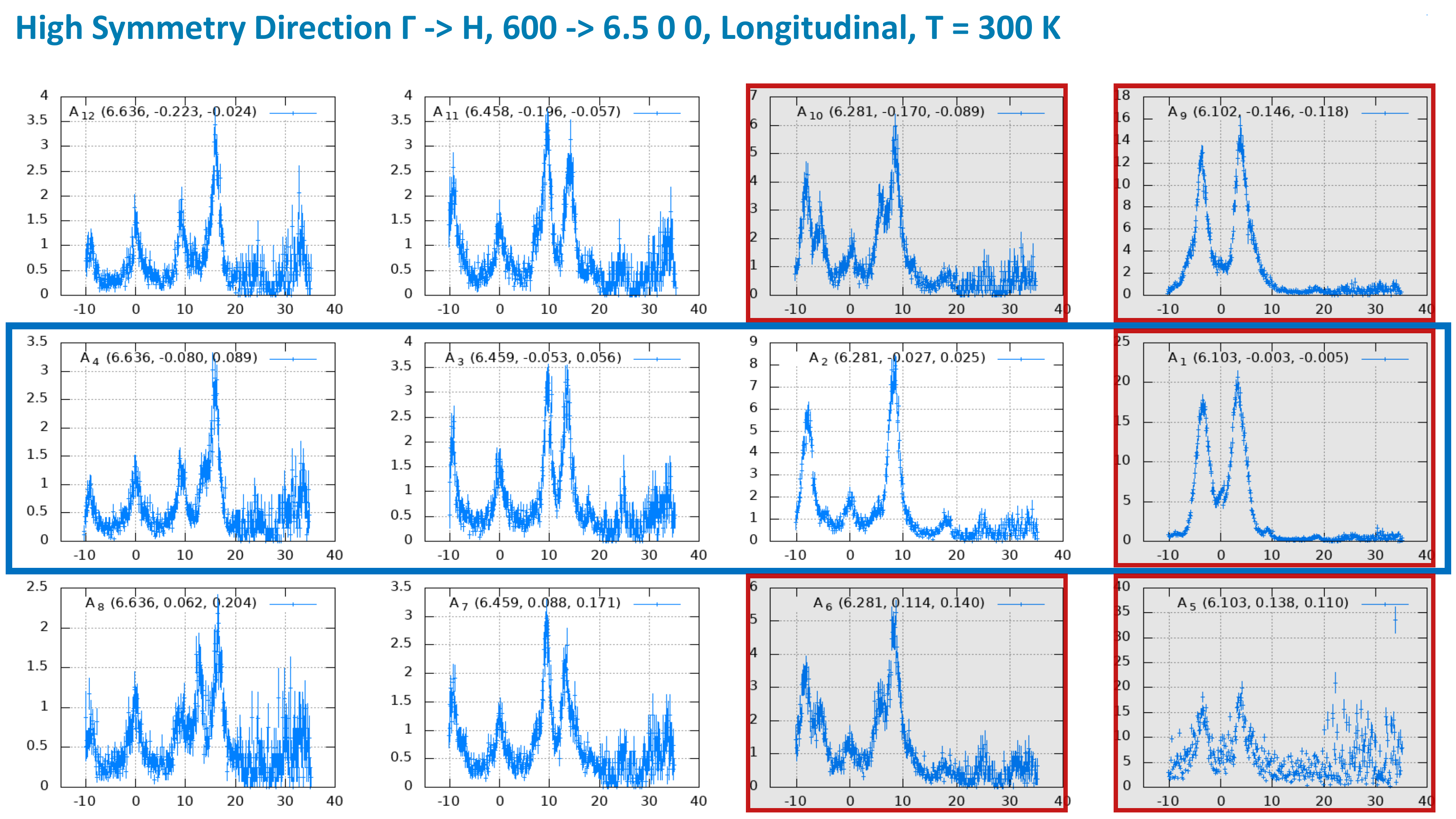}
\caption{Recorded spectra at 2, 20, and 300~K from top to bottom.
Constant \textbf{$Q$} scans were set up for high symmetry direction G$\rightarrow$H and longitudinal polarisaton of acoustic phonons highlighted by the blue-framed data set.
\textbf{$Q$} numbers are reported in the figures.
Red-framed spectra were not considered for the generalized S$(\omega)$ analysis.
\label{fig_SI_GHL1}}
\end{center}
\end{figure*}

\begin{figure*}[]
\begin{center}
\includegraphics[angle=0,width=125mm]{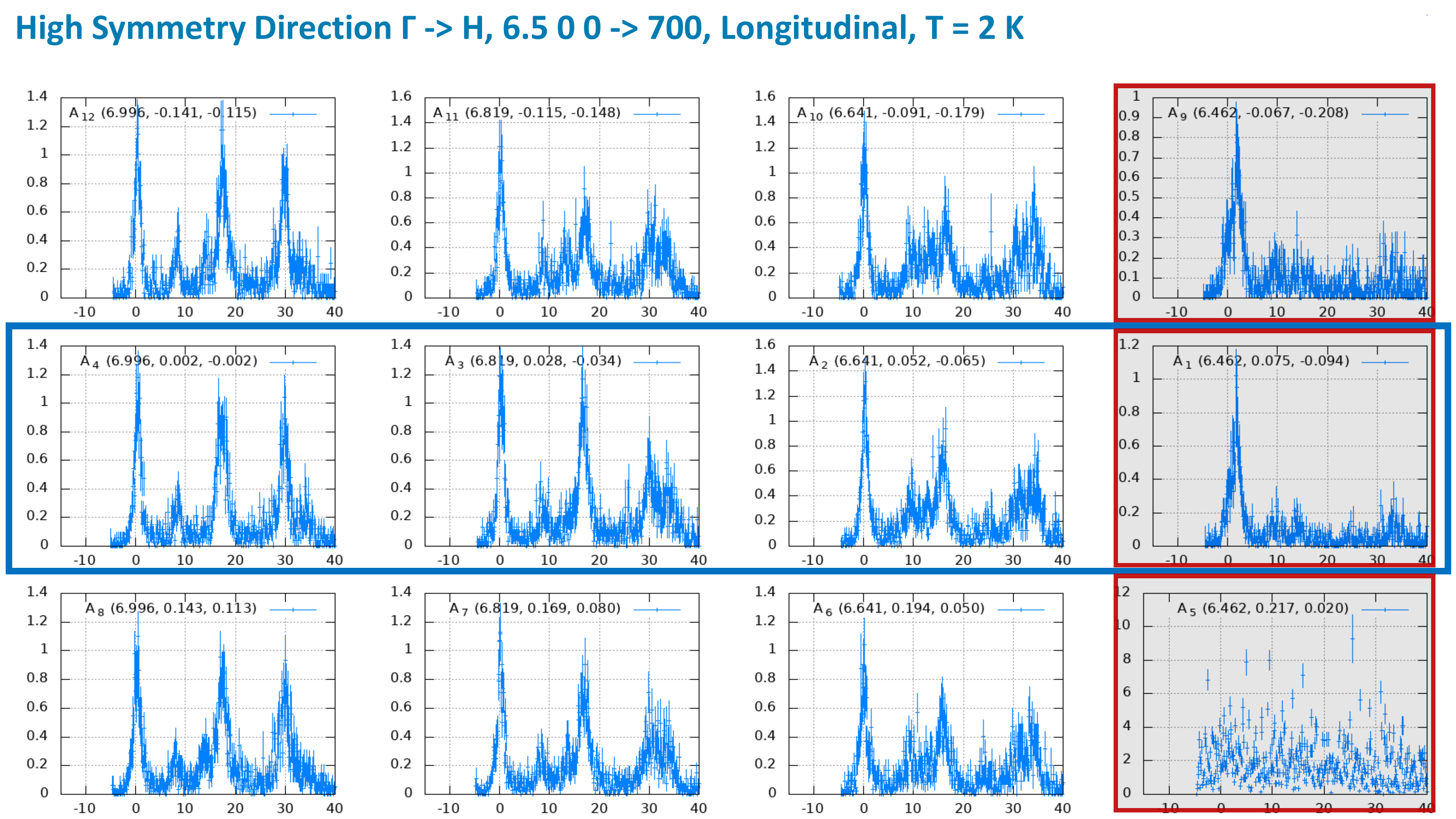}
\includegraphics[angle=0,width=125mm]{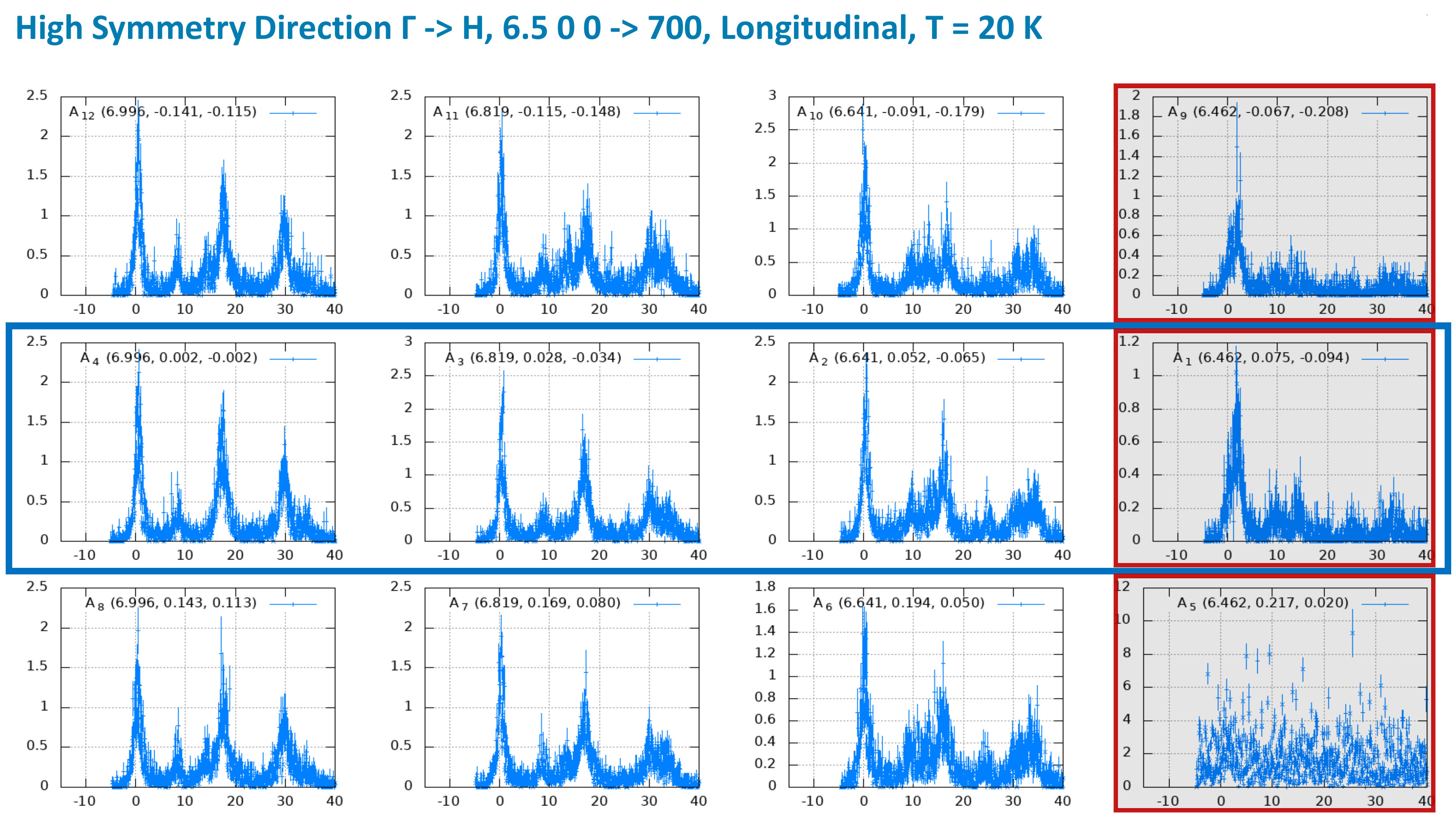}
\includegraphics[angle=0,width=125mm]{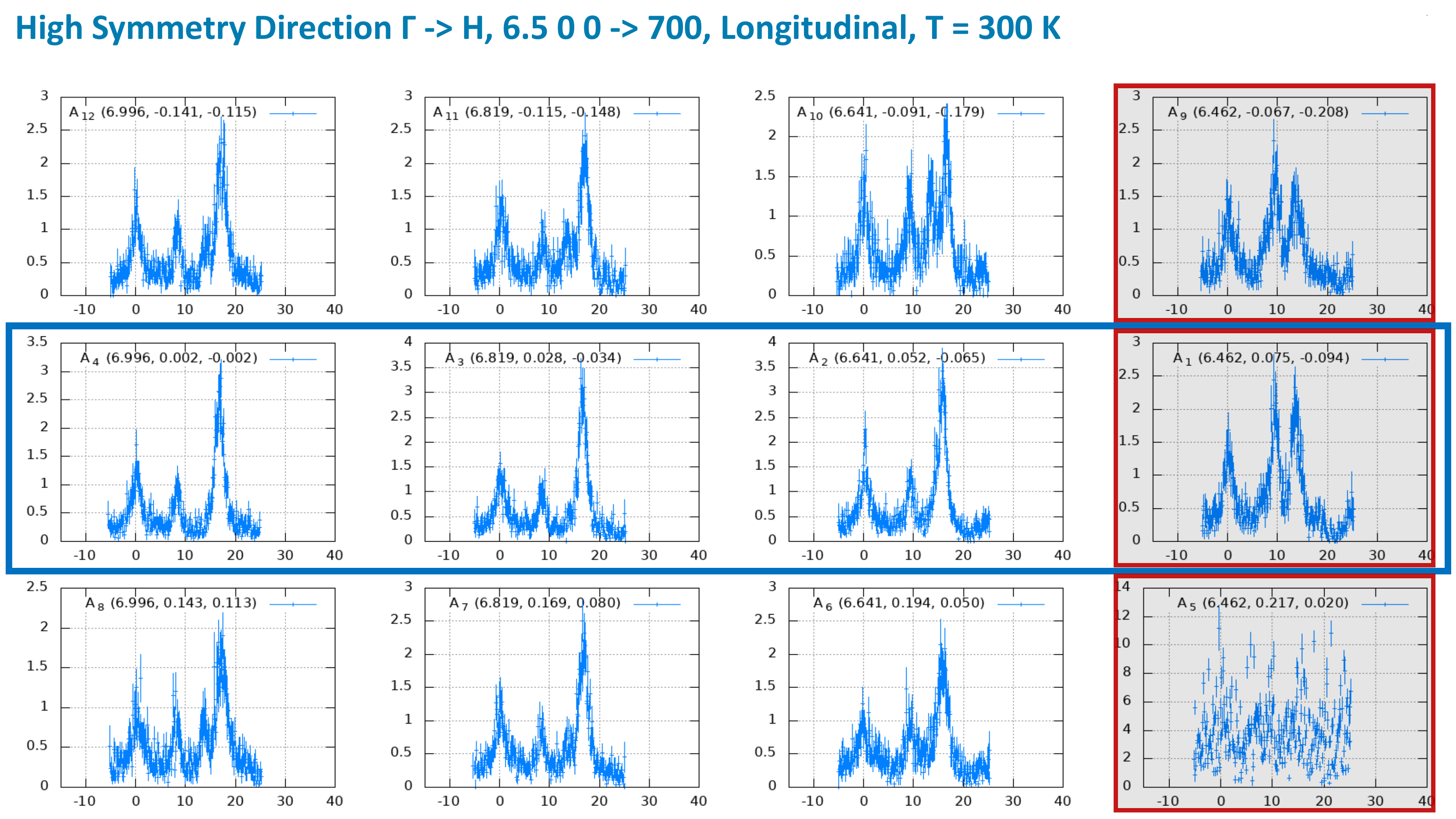}
\caption{Recorded spectra at 2, 20, and 300~K from top to bottom.
Constant \textbf{$Q$} scans were set up for high symmetry direction G$\rightarrow$H and longitudinal polarisaton of acoustic phonons highlighted by the blue-framed data set.
\textbf{$Q$} numbers are reported in the figures.
Red-framed spectra were not considered for the generalized S$(\omega)$ analysis.
\label{fig_SI_GHL2}}
\end{center}
\end{figure*}

\begin{figure*}[]
\begin{center}
\includegraphics[angle=0,width=125mm]{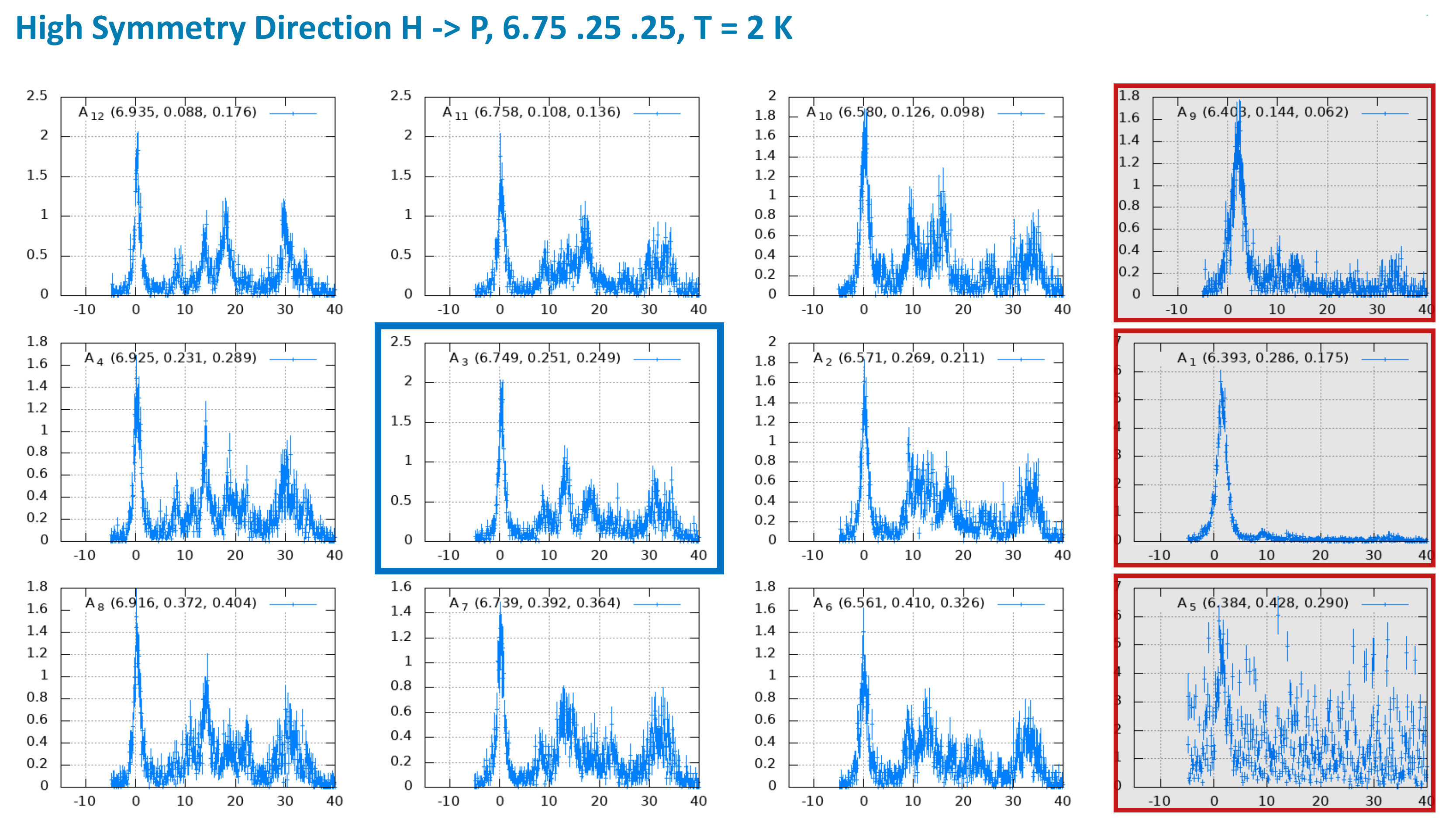}
\includegraphics[angle=0,width=125mm]{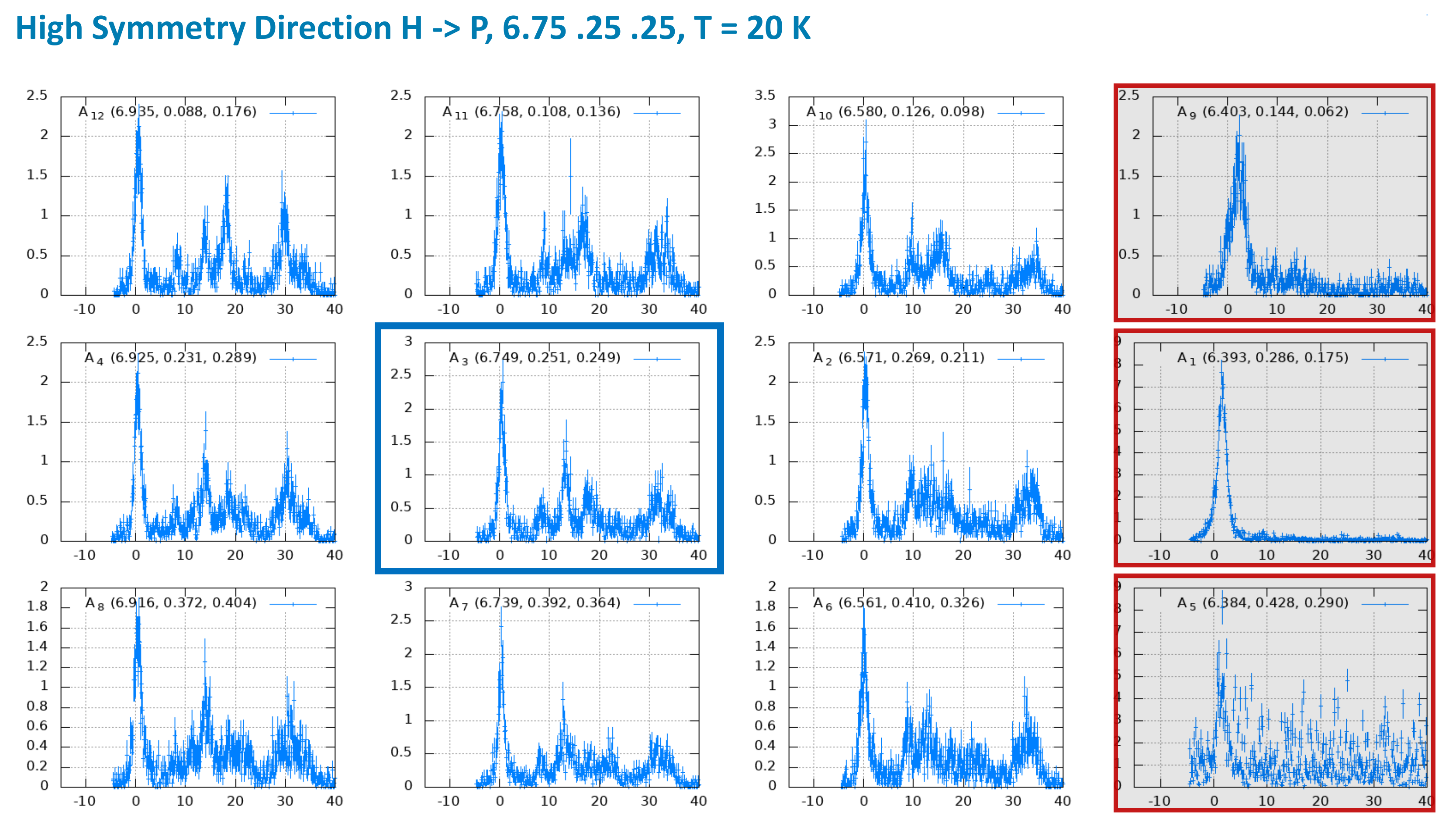}
\includegraphics[angle=0,width=125mm]{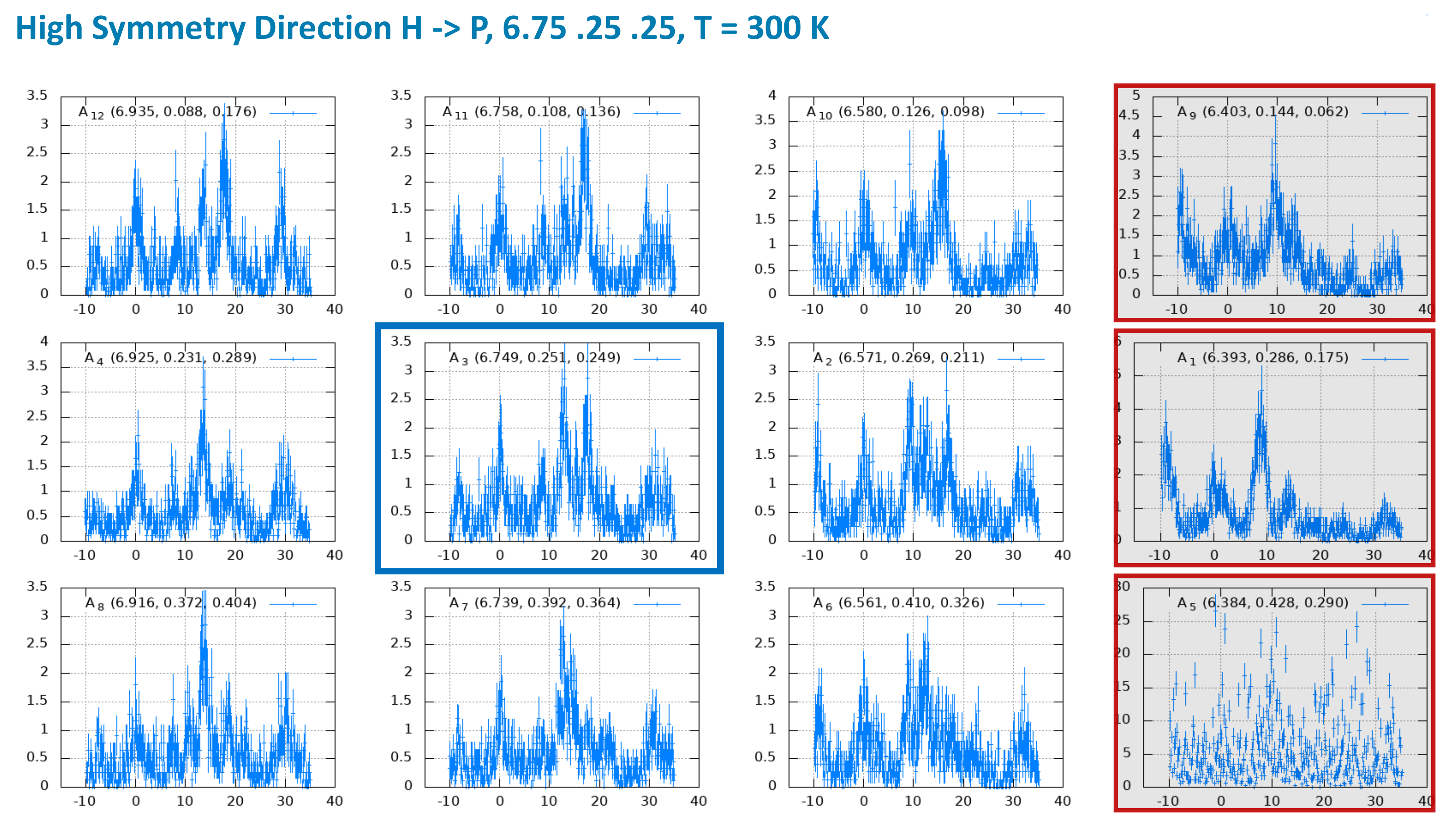}
\caption{Recorded spectra at 2, 20, and 300~K from top to bottom.
Constant ${\bf Q}$ scans were set up for the mid point at high symmetry direction H$\rightarrow$P highlighted by the blue-framed data set.
\textbf{$Q$} numbers are reported in the figures.
Red-framed spectra were not considered for the generalized S$(\omega)$ analysis.
\label{fig_SI_HPM}}
\end{center}
\end{figure*}

\begin{figure*}[]
\begin{center}
\includegraphics[angle=0,width=125mm]{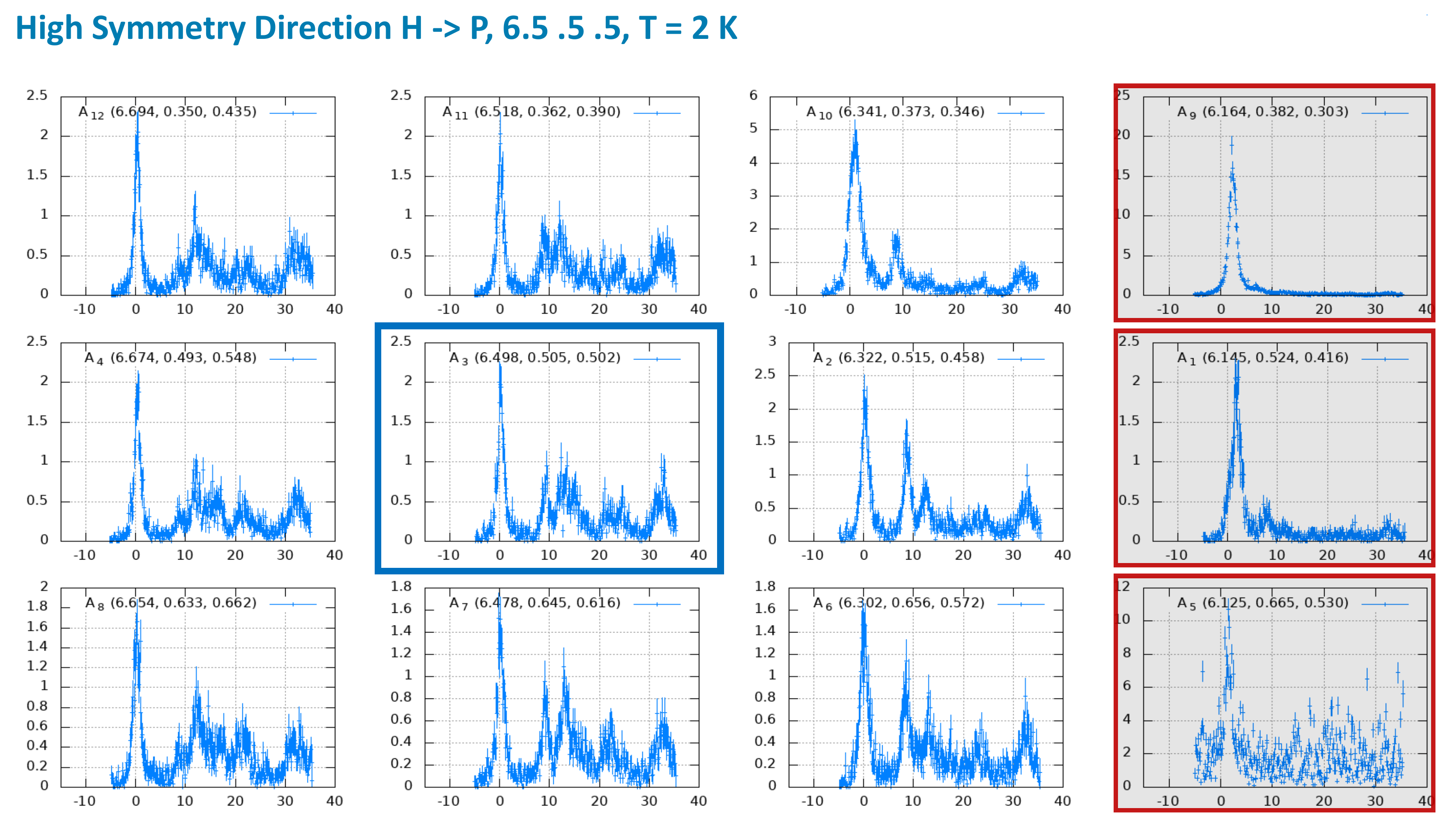}
\includegraphics[angle=0,width=125mm]{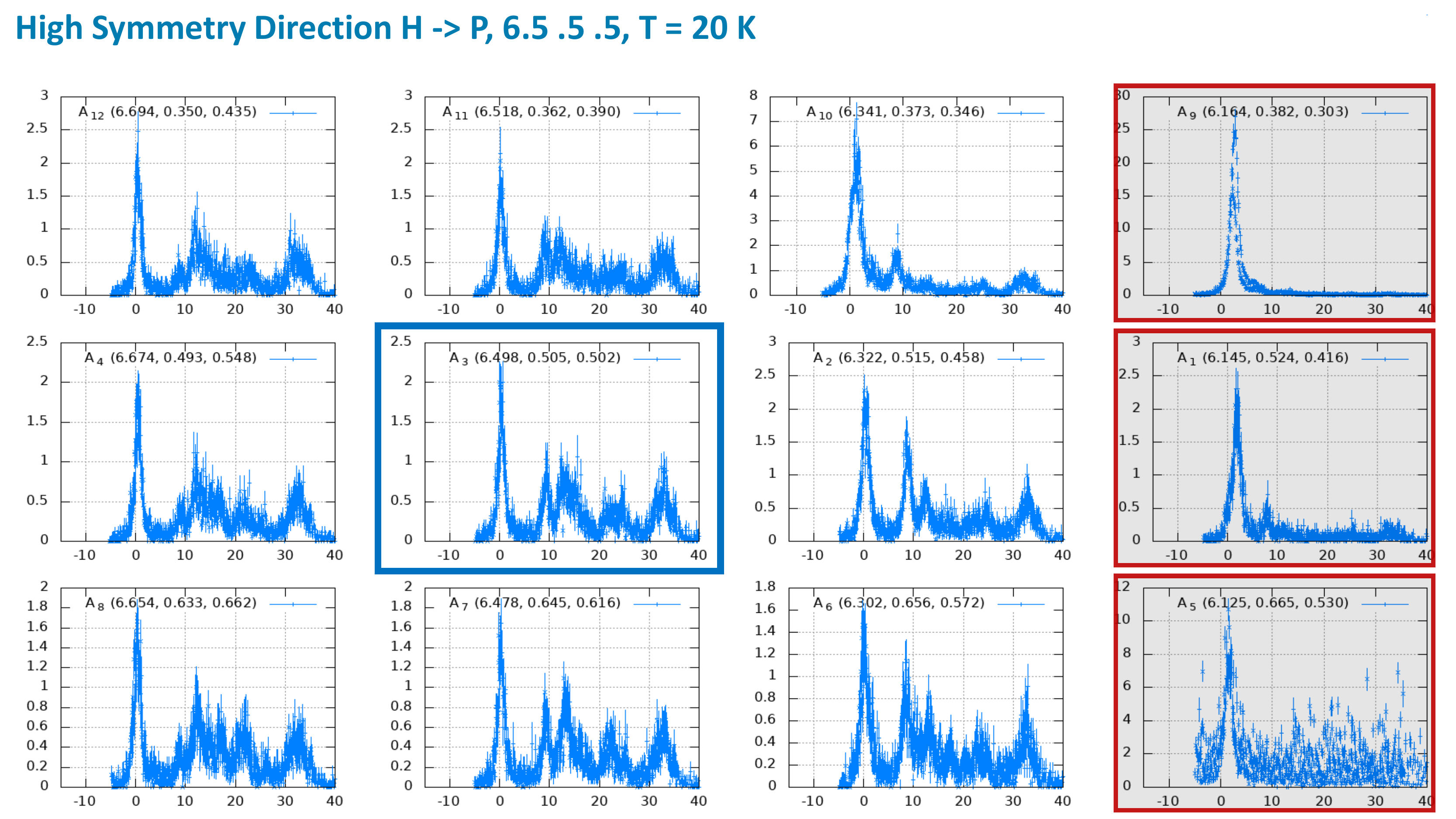}
\includegraphics[angle=0,width=125mm]{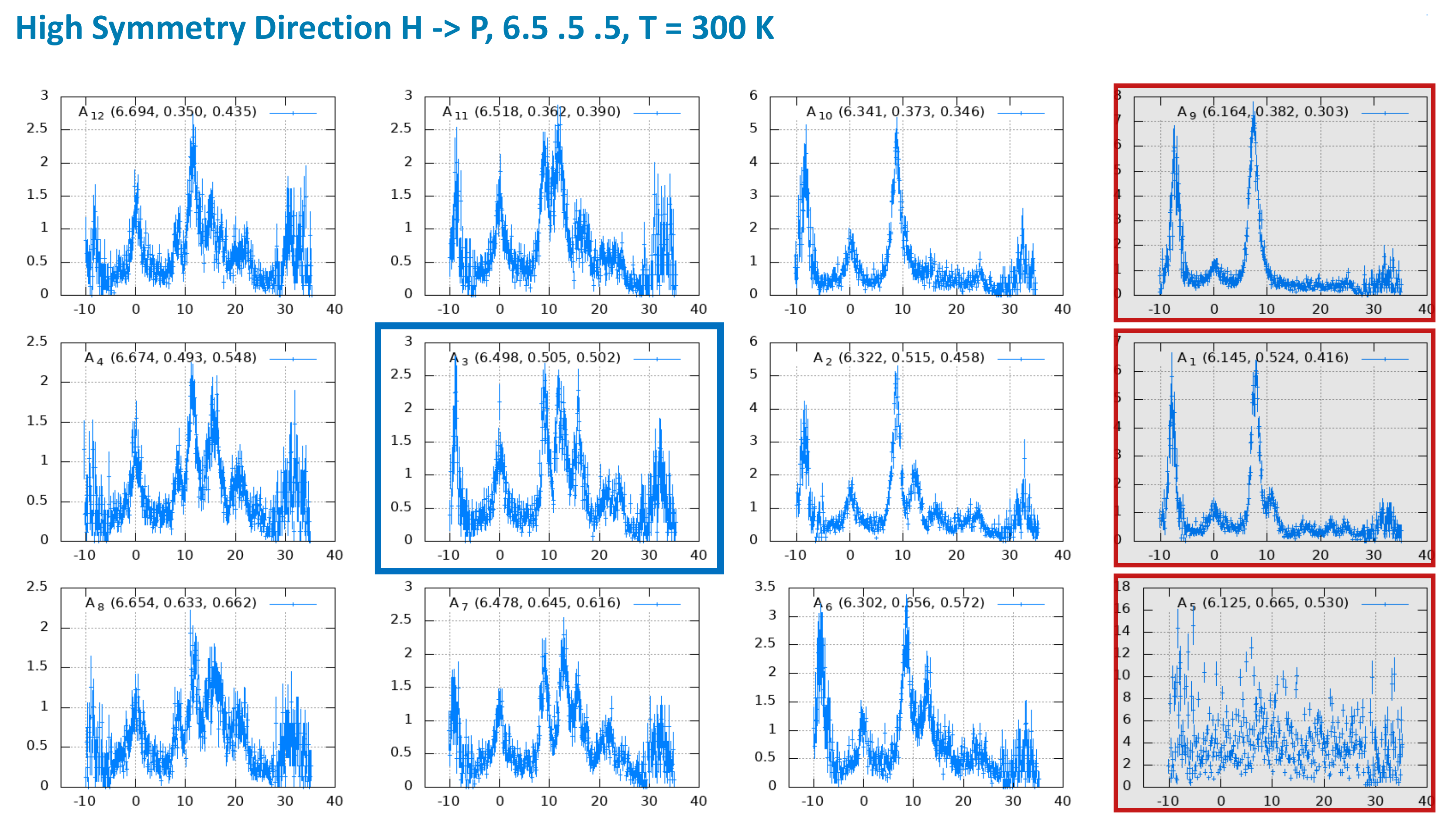}
\caption{Recorded spectra at 2, 20, and 300~K from top to bottom.
Constant ${\bf Q}$ scans were set up for the high symmetry point P highlighted by the blue-framed data set.
${\bf Q}$ numbers are reported in the figures.
Red-framed spectra were not considered for the generalized S$(\omega)$ analysis.
\label{fig_SI_HPE}}
\end{center}
\end{figure*}

Here we present in a concise form the inelastic x--ray (IXS) data and their analysis discussed in the main text. All recorded IXS scans are reported in Figs.~\ref{fig_SI_GNT1}-\ref{fig_SI_HPE}.
Recorded intensities were corrected for different efficiencies of the detector units.
The setup of the analyzers was chosen to follow the longitudinal or transverse polarization of acoustic modes along the high symmetry directions..
This was successfully carried out for 3 (transverse polarization) and 4 (longitudinal polarization) out of 12 analyzers with a single setup of the multi-analyzer unit.
All S$(\textbf{Q},\omega)$ spectra along high symmetry directions are highlighted by blue-framed boxes in the figures below.
Only these S$(\textbf{Q},\omega)$ spectra were taken into account for the evaluation of the phonon dispersion relation discussed in and reported in Fig.~6 of the main text.

The estimation of the temperature effect on the position of the phonon modes was carried out on merged spectra resulting in a generalized signal S$(\omega)$ of improved statistics.
Spectra with highly dispersive modes and strongly asymmetric elastic lines were not considered for the computation of S$(\omega)$. Those spectra are marked by red--framed boxes in the figures below.
Note that none of the spectra from the orientation G$\rightarrow$N (600$\rightarrow$6 .5 .5) shown in Fig.~\ref{fig_SI_GNT1} was taken into consideration for a generalized signal due to the enhanced mode dispersion.

Two evaluation approaches were attempted.
On one hand the spectra were merged as recorded.
On the other the energy axis of the single S$(\textbf{Q},\omega)$ spectra was adapted to center the elastic line exactly at 0 before merging.
The second approach was carried out to account for a hypothetical misalignment of the specimen during scans. 
Note that both data treatment strategies lead to congruent results with the merged-as-recorded data showing a stronger effect in favor of the data interpretation presented in the main text.
For this reason we only focus on and show the adapted-center-line data in the main text and this document.

\begin{figure*}[]
\begin{center}
\includegraphics[angle=0,width=120mm]{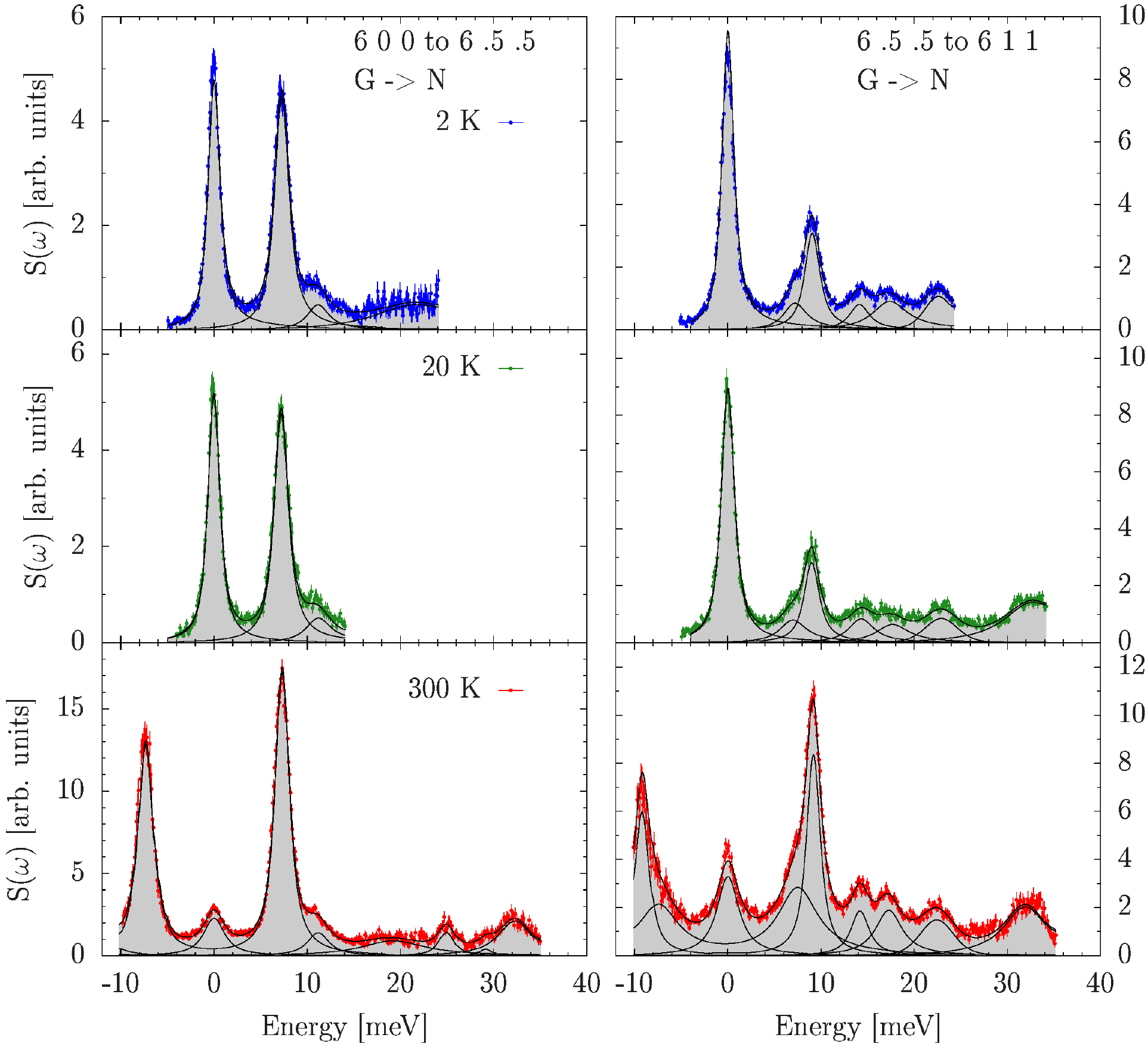}
\includegraphics[angle=0,width=120mm]{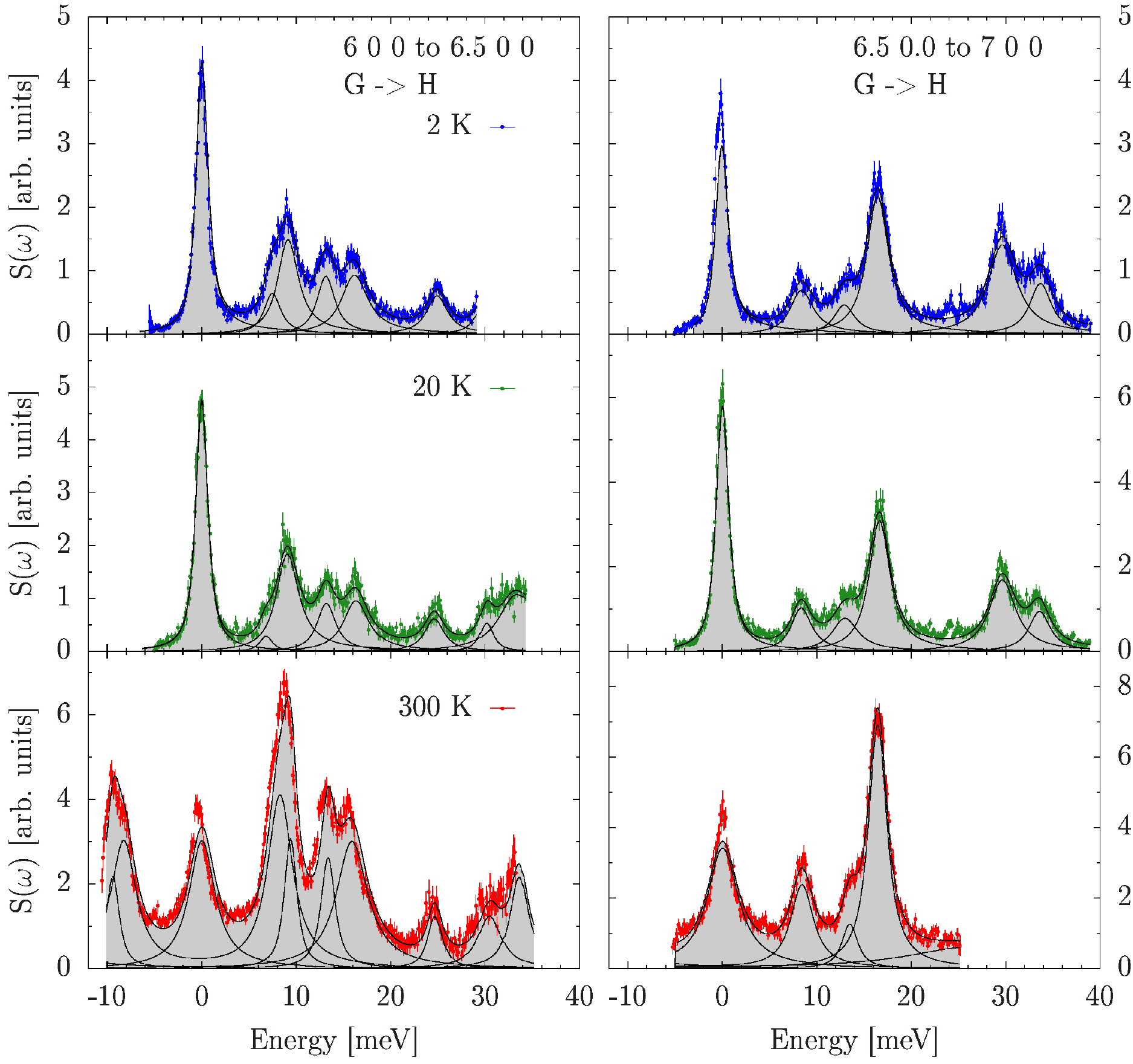}
\caption{Lorentzian fits convoluted with instrument resolution to the generalized spectra S$(\omega)$ for the multi-analyzer setup for direction G$\rightarrow$N (top) and G$\rightarrow$H (bottom). Temperatures are 2 (blue), 20 (green), and 300~K (red) from top to bottom of the corresponding setup.
\label{fig_gnt_ghl}}
\end{center}
\end{figure*}

The final evaluation step comprised a fitting of the generalized S$(\omega)$ with a set of Lorentzians whose number was adapted to the number of obvious phonon peaks.
Resolution function was taken into account by a convolution procedure.
Thereby an analytical pseudo-Voigt function with 1.45~meV full width at half maximum and an 80~\% Lorentzian contribution was chosen.
Those parameters represent the average over the different resolutions of the single analyzers measured and freely fitted with pseudo-Voigt lines. 
The approximated signals are reported with the generalized S$(\omega)$ in Fig.~7 of the main text and Fig.~\ref{fig_gnt_ghl} of this document.
The derived energy parameters are listed in Table~\ref{tab_peak_positions}. 
%
%==========================================================================
%
%
%==========================================================================
%
%
%==========================================================================
%

\begin{table}
\caption{Peak position parameters from Lorentzian fits to the generalized signal S$(\omega)$.}
\vspace{0.5cm}
\resizebox{\linewidth}{!}{
\begin  {tabular}{|c|c|c|c| }\hline
Orientation  & 2~K & 20~K & 300~K \\  \hline
G$\rightarrow$N & 7.25(1) & 7.21(1) & 7.31(1)\\
600 to 6 .5 .5 & 11.09(6) & 10.98(7) & 11.15(6)\\ 
& -- & -- & 19.1(3)\\ 
& -- & -- & 24.94(7)\\ 
& -- & -- & 32.08(8)\\  \hline
G$\rightarrow$N & 7.04(5) & 6.8(1) & 7.41(6)\\
6 .5 .5 to 611 & 9.00(2) & 8.98(3) & 9.18(1)\\
& 14.12(5) & 14.25(9) & 14.19(4)\\
& 17.23(5) & 17.5(2) & 17.23(5)\\
& 22.49(4) & 22.84(8) & 22.35(5)\\
& -- & 31.9(1) & 31.4(1)\\
& -- & 33.6(1) & 33.19(9)\\ \hline
G$\rightarrow$H & 7.49(5) & 6.8(2) &  8.26(5)\\ 
600 to 6.5 00 & 9.06(3) & 9.01(6) & 9.40(3)\\ 
& 13.13(4) & 13.17(7) & 13.37(3)\\ 
& 16.13(4) & 16.28(9) & 15.89(6)\\ 
& 24.88(4) & 24.60(7) & 24.67(7)\\ 
& 29.77(3) & 30.18(9) & 30.5(1)\\ 
& -- & 33.2(1) & 33.62(6)\\  \hline
G$\rightarrow$H  & 8.20(4) & 8.30(5) & 8.42(1) \\
6.5 00 to 700 & 12.92(7) & 12.92(9) & 13.37(5) \\
& 16.45(2) & 16.61(3) & 16.43(1) \\
& 29.60(4) & 29.61(4) & -- \\
& 33.71(4) & 33.56(6) & -- \\ \hline
H$\rightarrow$P & 8.80(5) & 8.52(6) & 8.67(6)\\
6.75 .25 .25 & 13.26(5) & 13.19(4) & 13.29(6)\\
& 17.48(7) & 17.29(7) & 17.35(7)\\
& 22.3(2) & 21.9(2) & 22.2(2)\\ 
& 30.53(8) & 30.32(8) & 30.40(9)\\ 
& 33.80(7) & 33.63(6) & 33.5(1)\\  \hline
H$\rightarrow$P  & 8.54(1) & 8.63(3) & 8.57(2)\\
6.5 .5 .5 & 12.18(3) & 12.27(5) & 11.85(3)\\ 
& 15.87(7) & 16.1(1) & 15.77(5)\\ 
& 20.35(1) & 20.7(1) & 20.24(8) \\ 
& 22.19(4) & 22.2(1) & 21.98(7)\\ 
& 24.44(4) & 24.54(9) & 24.12(6)\\ 
& 30.73(9) & 31.00(9) & 30.8(2)\\ 
& 32.34(3) & 32.84(6) & 32.21(1)\\ \hline
\end{tabular}}
\label{tab_peak_positions}
\end{table}


\begin{thebibliography}{99}
 
\bibitem{Book1} R.E. Baumbach and M.B. Maple, Filled Skutterudites: Magnetic and Electrical Transport Properties, in Encyclopedia of Materials: Science and Technology Reference Work (2001), Editors, K.H. Jürgen Buschow et al, Elsevier Ltd.

\bibitem{Sales1996} B.C.Sales, D. Mandrus and R.K.Williams, Science {\bf 272},1325 (1996). 

\bibitem{Keppens1998} V. Keppens, D. Mandrus, B. C. Sales, B. C. Chakoumakos, P. Dai, R. Coldea, M. B. Maple, D. A. Gajewski, E. J. Freeman, and S. Bennington, Nature (London) {\bf 395}, 876 (1998).

\bibitem{EDBauer2002} E. D. Bauer, N. A. Frederick, P.-C. Ho, V. S. Zapf, and M. B. Maple, Phys. Rev. B {\bf 65}, 100506(R) (2002).

\bibitem{Kotegawa2003} H. Kotegawa, M. Yogi, Y. Imamura, Y. Kawasaki, G.-q. Zheng, Y. Kitaoka, S. Ohsaki, H. Sugawara, Y. Aoki, and H. Sato, Phys. Rev. Lett. {\bf 90}, 027001 (2003).

\bibitem {Adroja2005} D. T. Adroja, A. D. Hillier, J.-G. Park, E. A. Goremychkin, K. A. McEwen, N. Takeda, R. Osborn, B. D. Rainford, and R. M. Ibberson, Phys. Rev. B {\bf 72}, 184503 (2005).


\bibitem{Hachitani2006} K. Hachitani, H. Fukazawa, Y. Kohori, I. Watanabe, C. Sekine, and I. Shirotani, Phys. Rev. B {\bf 73}, 052408 (2006).

\bibitem {Curnoe2002} S.H. Curnoe, H. Harima, K. Takegahara, K. Ueda, Physica B {\bf 312-313}, 837 (2002).

\bibitem{Sugawara2002} H. Sugawara, T. D. Matsuda, K. Abe, Y. Aoki, H. Sato, S. Nojiri, Y. Inada, R. Settai, and Y. Onuki, Phys. Rev. B {\bf 66}, 134411 (2002). 

\bibitem{Iwasa2002} K. Iwasa, Y. Watanabe, K. Kuwahara, M. Kohgi, H. Sugawara, T. D. Matsuda, Y. Aoki, and H. Sato, Physica B {\bf 312-313}, 834 (2002).

\bibitem{Kohgi} M. Kohgi et al. J. Phys. Soc. Jpn. {\bf 72}, 1002  (2003).

\bibitem {Takeda2000} N. Takeda and Y. Ishikawa, J. Phys. Soc. Jpn. {\bf 69}, 868 (2000).

\bibitem{CeRu4Sb12_1} See, for example, Proc. Int. Conf. New Quantum Phenomena in Skutterudite and Related Systems, J. Phys. Soc. Jpn. {\bf 77} (2008) Suppl. A


\bibitem{Adroja2003} D. T. Adroja, J.-G. Park, K. A. McEwen, N. Takeda, M. Ishikawa, and J.-Y. So, Phys. Rev. B {\bf 68}, 094425 (2003).

\bibitem {Adroja2007} D. T. Adroja, J.-G. Park, E. A. Goremychkin, K. A. McEwen, N. Takeda, B. D. Rainford, K. S. Knight, J. W. Taylor, Jeongmi Park, H. C. Walker, R. Osborn, and P. S. Riseborough, Phys. Rev. B {\bf 75}, 014418 (2007).

\bibitem{Baumbach2008} R. E. Baumbach, P. C. Ho, T. A. Sayles, M. B. Maple, R. Wawryk, T. Cichorek, A. Pietraszko, and Z. Henkie, Proc. Natl. Acad. Sci. U.S.A {\bf 105}, 17307 (2008).

\bibitem{Sanada2004} S. Sanada, Y. Aoki, H. Aoki, A. Tsuchiya, D. Kikuchi, H. Sugawaray, and H. Sato, J. Phys. Soc. Jpn, {\bf 74}, 246 (2005). 

\bibitem{Sekine} C. Sekine, T. Uchiumi, I. Shirotani, and T. Yagi, Phys. Rev. Lett. {\bf 79}, 3218 (1997).

\bibitem{5} L. Bochenek, R. Wawryk, Z. Henkie, and T. Cichorek, Phys.
Rev. B {\bf 86}, 060511(R) (2012).

\bibitem{6} R. E. Baumbach, P. C. Ho, T. A. Sayles, M. B. Maple, R.
Wawryk, T. Cichorek, A. Pietraszko, and Z. Henkie, J. Phys.:
Condens. Matter {\bf 20}, 075110 (2008).

\bibitem{7}T. A. Sayles, R. E. Baumbach, W. M. Yuhasz, M. B. Maple, L.
Bochenek, R. Wawryk, T. Cichorek, A. Pietraszko, Z. Henkie,
and P.-C. Ho, Phys. Rev. B {\bf 82}, 104513 (2010).

\bibitem{8} A. Rudenko, Z. Henkie, and T. Cichorek, Solid State Commun.
{\bf 242}, 21 (2016).

\bibitem{Seyfarth} G. Seyfarth, J. P. Brison, M.-A. Méasson, J. Flouquet, K. Izawa, Y. Matsuda, H. Sugawara, and H. Sato, Phys. Rev. Lett. {\bf 95}, 107004 (2005).

\bibitem{Jeitschko} W. Jeitschko and D. Braun, Acta Crystallogr. Sect. B {\bf 33}, 3401 (1977).

\bibitem{Sato} H. Sato, H. Sugawara, Y. Aoki, and H. Harima, in Handbook of Magnetic Materials, edited by K. H. J. Buschow (North-Holland, Amsterdam, 2009), Vol. 18, p. 1.


\bibitem{Tarantini} C. Tarantini, A. Gurevich, J. Jaroszynski, F. Balakirev, E. Bellingeri, I. Pallecchi, C. Ferdeghini, B. Shen, H. H. Wen, and D. C. Larbalestier, Phys. Rev. B {\bf 84}, 184522 (2011).

\bibitem{Ren} C. Ren, Z.-S. Wang, H.-Q. Luo, H. Yang, L. Shan, and H.-H.Wen, Phys. Rev. Lett. {\bf 101}, 257006 (2008).

%\bibitem{Shi} X. Shi, J. Yang, L. Wu, J. R. Salvador, C. Zhang, W. L. Villaire, D. Haddad, J. Yang, Y. Zhu and Q. Li, Scientific Reports {\bf 5}, 14641 (2015).

%\bibitem{Bauer2002} E. D. Bauer, N. A. Frederick, P.-C. Ho, V. S. Zapf, and M. B. Maple, Phys. Rev. B {\bf 65}, 100506(R) (2012).

%\bibitem{Measson} M.-A. Measson, D. Braithwaite, J. Flouquet, G. Seyfarth, J. P. Brison, E. Lhotel, C. Paulsen, H. Sugawara, and H. Sato, Phys. Rev. B {\bf 70}, 064516 (2004).

%\bibitem{MacLaughlin} D. E. MacLaughlin, J. E. Sonier, R. H. Heffner, O. O. Bernal, Ben-Li Young, M. S. Rose, G. D. Morris, E. D. Bauer, T. D. Do, and M. B. Maple, Phys. Rev. Lett. {\bf 89}, 157001 (2002). 

%\bibitem{Aoki} Y. Aoki, A. Tsuchiya, T. Kanayama, S. R. Saha, H. Sugawara, H. Sato, W. Higemoto, A. Koda, K. Ohishi, K. Nishiyama, and R. Kadono, Phys. Rev. Lett. {\bf 91}, 067003 (2003).







\bibitem{Kasahara} Y. Kasahara, T. Iwasawa, H. Shishido, T. Shibauchi, K. Behnia, Y. Haga, T. D. Matsuda, Y. Onuki, M. Sigrist, and Y. Matsuda, Phys. Rev. Lett. {\bf 99}, 116402 (2007).

\bibitem{Hill} R.W. Hill, S. Li, M. B. Maple, and L. Taillefer, Phys. Rev. Lett. {\bf 101}, 237005 (2008).

\bibitem{Kittaka} S. Kittaka, Y. Aoki, Y. Shimura, T. Sakakibara, S. Seiro,
C. Geibel, F. Steglich, H. Ikeda, and K. Machida, Phys. Rev.
Lett. {\bf 112}, 067002 (2014). 

\bibitem{Hor} Y. S. Hor, A. J. Williams, J. G. Checkelsky, P. Roushan, J. Seo, Q. Xu, H.W. Zandbergen, A. Yazdani, N. P. Ong, and R. J. Cava, Phys. Rev. Lett. {\bf 104}, 057001 (2010).

\bibitem{Zhang} J. L. Zhang, S. J. Zhang, H. M. Weng, W. Zhang, L. X. Yang, Q. Q. Liu, S. M. Feng, X. C. Wang, R. C. Yu, L. Z. Cao, L.Wang, W. G. Yang, H. Z. Liu, W. Y. Zhao, S. C. Zhang, X. Dai, Z. Fang, and C. Q. Jin, Proc. Natl. Acad. Sci. U.S.A. {\bf 108}, 24 (2011).



\bibitem{Namiki} T. Namiki, C. Sekine, K. Matsuhira, M. Wakeshima, and I.
Shirotani, J. Phys. Soc. Jpn. {\bf 77}, 336 (2008).



\bibitem{Shirotani} I. Shirotani, K. Ohno, C. Sekine, T. Yagi, T. Kawakami, T. Nakanishi, H. Takahashi, J. Tang, A. Matsushitad, T. Matsumotod, Phys. B: Cond. Mat. {\bf 281-282}, 1021-1023 (2000). 

\bibitem{Shirotani1997} I. Shirotani, T. Uchiumi, K. Ohno, C. Sekine, Y. Nakazawa, K. Kanoda, S. Todo, and T. Yagi, Phys. Rev. B {\bf 56}, 7866 (1997).  

%\bibitem{Shirotani2005} I. Shirotani, S. Sato, C. Sekine, K. Takeda, I. Inagawa and T. Yagi, J. Phys.: Condens. Matter {\bf 17}, 7353 (2005).

\bibitem{Klotz} J. Klotz, K. G\"{o}tze, V. Lorenz, Yu. Prots, H. Rosner, H. Harima, L. Bochenek, Z. Henkie, T. Cichorek, I. Sheikin, and J. Wosnitza, Phys. Rev. B {\bf 100}, 205106 (2019). 

\bibitem{Mizukami11} Y. Mizukami, M. Kończykowski, O. Tanaka, J. Juraszek, Z. Henkie, T. Cichorek, and T. Shibauchi, Phys. Rev. Research {\bf 2}, 043428 (2020).

\bibitem{Bochenek} L. Bochenek, R. Wawryk, Z. Henkie, and T. Cichorek, Phys. Rev. B, {\bf 86},060511 (2012).

\bibitem{Henkie} Z. Henkie, M. B. Maple, A. Pietraszko, R. Wawryk, T. Cichorek, R. E. Baumbach, W. M. Yuhasz, and P.-C. Ho, J. Phys. Soc. Jpn. {\bf 77}, 128-134 (2008). 



\bibitem{Pratt2000} F. L. Pratt, Physica B {\bf 289-290}, 710 (2000).

\bibitem{BL35XU-Baron-JPCS-2000} A. Q. R. Baron, Y. Tanaka, S. Goto, K. Takeshita, T. Matsushita, T. Ishikawa, An X-ray scattering beamline for studying dynamics, Journal of Physics and Chemistry of Solids, {\bf 61}, 461-465 (2000).

\bibitem{VASP-Kresse-1996} G. Kresse and J. Furthm\"{u}ller, Comput. Mater. Sci. {\bf 6}, 15 (1996).

\bibitem{PAW-Kresse-1999} G. Kresse and D. Joubert, Phys. Rev. {\bf 59} , 1758 (1999).

\bibitem{GGA-PBE-PRL-1996} J. P. Perdew, K. Burke, and M. Ernzerhof, Phys. Rev. Lett. {\bf 77}, 3865 (1996).

\bibitem{Monkhorst-PRB-1976} H. J. Monkhorst and J. D. Pack, Phys. Rev. B {\bf 13}, 5188 (1976).

\bibitem{Phonon-Parlinski-1999} K. Parlinski, AIP Conf. Proc. {\bf 479}, 121 (1999).

\bibitem{Juraszek2020} J. Juraszek, R. Wawryk, Z. Henkie, M. Konczykowski, and T. Cichorek, Phys. Rev. Lett. {\bf 124}, 027001 (2020). 

\bibitem{Rainford}B.D. Rainford et al. Hyperfine Interactions {\bf 87} 1129-1134 (1994).

\bibitem{Bhattacharyya1} A. Bhattacharyya, M. R. Lees, K Panda, P. P. Ferreira, T. T. Dorini, Emilie Gaudry, L. T. F. Eleno, V. K. Anand, J. Sannigrahi, P. K. Biswas, R. Tripathi, D. T. Adroja, Phys. Rev. Matt. {\bf 6}, 064802, (2022).


\bibitem{Bhattacharyyarev} A. Bhattacharyya, D. T. Adroja, M. Smidman, and V. K. Anand, Sci. China-Phys. Mech. Astron. {\bf 61}, 127402 (2018).

\bibitem{ThCoC2} A. Bhattacharyya, D. T. Adroja, K. Panda, Surabhi Saha, Tanmoy Das, A. J. S. Machado, T. W. Grant, Z. Fisk, A. D. Hillier, and P. Manfrinetti, Phys. Rev. Lett. {\bf 122}, 147001 (2019).

\bibitem{Panda} K. Panda, A. Bhattacharyya, D. T. Adroja, N. Kase, P. K. Biswas, S. Saha, T. Das, M. Lees, A. D. Hillier, Phys. Rev. B {\bf 99}, 174513 (2019).

\bibitem{Sonier} J. E. Sonier, J. H. Brewer, and R. F. Kiefl, Rev. Mod. Phys. {\bf 72}, 769 (2000).

\bibitem{Chia} E. E. M Chia, Elbert. S.M. Salamon, H. Sugawara and H. Sato, Phys. Rev. B, {\bf 69}, 180509 (R) (2004).
\bibitem{Prozorov} R. Prozorov and R. W. Giannetta, Magnetic penetration depth in
unconventional superconductors, Supercond. Sci. Technol., {\bf19}, R41 (2006).

\bibitem {Annett} J. F. Annett, Adv. Phys. {\bf 39}, 83 (1990).

\bibitem{Pang} G. M. Pang, M. Smidman, W. B. Jiang, J. K. Bao, Z. F. Weng, Y. F. Wang, L. Jiao, J. L. Zhang, G. H. Cao, and H. Q. Yuan, Phys. Rev. B {\bf 91}, 220502(R) (2015).

\bibitem{Khasanov2009} R. Khasanov, D. V. Evtushinsky, A. Amato, H.-H. Klauss, H. Luetkens, Ch. Niedermayer, B. Büchner, G. L. Sun, C. T. Lin, J. T. Park, D. S. Inosov, and V. Hinkov, Phys. Rev. Lett. {\bf 102}, 187005 (2009).
 
\bibitem{Avci} S. Avci, O. Chmaissem, D. Y. Chung, S. Rosenkranz, E. A. Goremychkin, J. P. Castellan, I. S. Todorov, J. A. Schlueter, H. Claus, A. Daoud-Aladine, D. D. Khalyavin, M. G. Kanatzidis, and R. Osborn, Phys. Rev. B {\bf 85}, 184507 (2012).

\bibitem{Khasanov2007} R. Khasanov, A. Shengelaya, A. Bussmann-Holder, H. Keller, Springer-Verlag Berlin Heidelberg, 177–190 (2007).

\bibitem{Biswas} P. K. Biswas, A. Amato, C. Baines, R. Khasanov, H. Luetkens, Hechang Lei, C. Petrovic, and E. Morenzoni, Phys. Rev. B {\bf 88}, 224515 (2013). 

\bibitem{HfIrSi} A. Bhattacharyya, K. Panda, D. T. Adroja, N. Kase, P. K. Biswas, S. Saha, T. Das, M. R. Lees, A. D. Hillier, J. Phys.: Cond. Mat. {\bf 32}, 085601 (2020).

\bibitem{CeIr3} D. T. Adroja, A. Bhattacharyya, Y. J. Sato, M. R. Lees, P. K. Biswas, K. Panda, V. K. Anand, Gavin B. G. Stenning, A. D. Hillier, and D. Aoki Phys. Rev. B {\bf 103}, 104514 (2021). 

\bibitem{Zr5Pt3} A. Bhattacharyya, P. P. Ferreira, K. Panda, F. B. Santos, D. T. Adroja, K. Yokoyama, T. T. Dorini, L. T. F. Eleno, A. J. S. Machado J. Phys.: Cond. Matt. {\bf 34}, 035602 (2022).

\bibitem{Koza-JPSJ-2013} M.M. Koza, D.A. Droja, N.T. Akeda, Z. Henkie, and T. Cichorek, J. Phys. Soc. Jpn. {\bf 82}, 114607 (2013).

\bibitem{Tutuncu-PRB-2017} T\"ut\"unc\"u, H.M., Karaca, Ertu\v{g}rul, Srivastava, G.P., Physical Review B {\bf 95}, 214514 (2017).

\bibitem{CeCo2Ga8} A. Bhattacharyya, D. T. Adroja, J. S. Lord, L. Wang, Y. Shi, K. Panda, H. Luo, and A. M. Strydom Phys. Rev. B {\bf 101}, 214437 (2020).

\bibitem{Tutuncu2017} H. M. T\"ut\"unc\"u, Ertu\ifmmode \check{g}\else \v{g}\fi{}rul Karaca and G.P. Srivastava, Phys. Rev. B {\bf 95}, 214514 (2017).

\end{thebibliography}
\end{document}